\documentclass{sig-alternate}

\usepackage{graphicx}
\usepackage{array,booktabs,ragged2e}
\usepackage[tight,footnotesize]{subfigure}

\usepackage{color}
\usepackage{amsmath}
\usepackage{amsfonts}
\usepackage{epsfig}
\usepackage{url}
\usepackage{multirow}
\usepackage[ruled,vlined,linesnumbered]{algorithm2e}
\usepackage{setspace}
\usepackage{wrapfig}
\usepackage{comment}
\usepackage{lipsum}

\newcommand{\z}{{\mathbf{z}}}
\newcommand{\x}{{\mathbf{x}}}
\newcommand{\y}{{\mathbf{y}}}

 \newtheorem{definition}{Definition}
\newcommand{\hide}[1]{}

\newcommand{\method}{{\sc StreamSpot}}
\newcommand{\simhash}{{\sc SimHash}}
\newcommand{\minhash}{{\sc MinHash}}
\newcommand{\streamhash}{{\sc StreamHash}}
\newcommand{\okbfs}{{\sc OkBFT}}
\newcommand{\DA}{{\sc YDC}}
\newcommand{\DB}{{\sc GFC}}
\newcommand{\DC}{{\sc ALL}}

\newcommand{\bit}{\begin{itemize}}
	\newcommand{\eit}{\end{itemize}}
\newcommand{\ben}{\begin{enumerate}}
	\newcommand{\een}{\end{enumerate}}
\newcommand{\beq}{\begin{equation}}
\newcommand{\eeq}{\end{equation}}

\title{Fast Memory-efficient Anomaly Detection in \\ Streaming Heterogeneous Graphs}

\numberofauthors{1}
\author{
\makebox[98pt]{Emaad A. Manzoor$^\star$ \quad Sadegh Momeni$^\dagger$\quad Venkat N. Venkatakrishnan$^\dagger$ \quad Leman Akoglu$^\star$}\\
\affaddr{$^\star$Stony Brook University}\\
\affaddr{$^\dagger$University of Illinois at Chicago}\\
\email{\{emanzoor, leman\}@cs.stonybrook.edu, \{smomen2,venkat\}@uic.edu}\\
}

\begin{document}

\maketitle

\begin{abstract}

Given a stream of heterogeneous graphs
containing different types of nodes and edges,
how can we spot anomalous ones in real-time while 
consuming bounded memory?
This problem is motivated by and generalizes from its
application in security to host-level advanced persistent threat (APT) detection.
We propose \method, a clustering based anomaly detection approach
that addresses challenges in two key fronts:
(1) {\em heterogeneity}, and (2) {\em streaming nature}.
We introduce a new similarity function
for heterogeneous graphs
that compares two graphs based on their
relative frequency of local substructures, represented as short strings.
This function lends itself to a vector representation of a graph, which is
($a$) fast to compute, and
($b$) amenable to a \emph{sketched} version with bounded size 
that preserves similarity.
\method~exhibits desirable properties that a streaming application
requires---it is
($i$) fully-streaming; processing the stream one edge at a time as it arrives,
($ii$) memory-efficient; requiring constant space for the sketches and the clustering,
($iii$) fast; taking constant time to update the graph sketches and the cluster summaries that can process
over $100K$ edges per second, and
($iv$) online; scoring and flagging anomalies in real time.
Experiments on datasets containing simulated system-call flow graphs from normal
browser activity and various attack scenarios (ground truth) show that our proposed
\method~is {\em high-performance}; achieving above $95$\% detection accuracy with small delay,
as well as competitive time and memory usage.

\end{abstract}

\section{Introduction}
\label{sec:intro}

%
%
%
%
%

Anomaly detection is a pressing problem for various critical tasks in security, finance, medicine, and so on.
In this work, we consider the anomaly detection problem
for streaming heterogeneous graphs, which contain different types of nodes and edges.
The input is a stream of timestamped and typed edges, where the source and detination nodes are also typed.
Moreover, multiple such graphs may be arriving over the stream simultaneously, that is, edges that belong to different graphs may be interleaved.
The goal is to accurately and quickly identify the anomalous graphs that are significantly different from what has been observed over the stream thus far, while meeting several important needs of the driving applications including fast real-time detection and bounded memory space usage. 

The driving application that motivated our work is the advanced persistent threat (APT) detection problem in security, although the above abstraction can appear in numerous other settings (e.g., software verification).
In the APT scenario, we are given a stream of logs capturing the events occuring in the system.
These logs are used to construct what is called \emph{information flow graphs}, in which edges depict data or control dependencies. Both the nodes and edges of the flow graphs are typed. Examples to node types are {\tt file}, {\tt process}, etc. and edge types include various system calls such as {\tt read}, {\tt write}, {\tt fork}, etc. as well as other parent-child relations. 
Within a system, an information flow corresponds to a unit of functionality (e.g., checking email, watching video, software updates, etc.). Moreover,
multiple information flows may be occurring in the system simultaneously.
The working assumption for APT detection is that the information flows induced by malicious activities in the system are
sufficiently different from the normal behavior of the system.
Ideally, the detection is to be done in real-time with small computational overhead and delay.
As the system-call level events occur rapidly in abundance, it is also crucial to process them
in memory while also incurring low space overhead.
The problem then can be cast as real-time anomaly detection in streaming heterogeneous graphs with bounded space and time, as stated earlier.

Graph-based anomaly detection has been studied in the past two decades.
Most work is for static homogeneous graphs \cite{DBLP:journals/datamine/AkogluTK15}.
Those for typed or attributed graphs aim to find deviations from frequent substructures \cite{conf/kdd/NobleC03,conf/icdm/EberleH07,conf/sdm/LiuYYHY05},
anomalous subgraphs \cite{Gupta2014,Perozzi2016}, and
community outliers  \cite{conf/kdd/GaoLFWSH10,Perozzi2014}, all of which are designed for static graphs.
For streaming graphs various techniques have been proposed for
clustering \cite{conf/sdm/AggarwalZY10} and connectivity anomalies \cite{conf/icde/AggarwalZY11} for plain graphs, which are recently extended to graphs with attributes \cite{conf/sdm/YuZ13,conf/sdm/McConville15}. (See Sec. \ref{sec:related})
Existing approaches are not, at least directly, applicable to our motivating scenario
as they do not exhibit all of the desired properties simultaneously; namely,
handling heterogeneous graphs, streaming nature,
low computational and space overhead, 
and real-time anomaly detection.


To address the problem for streaming heterogeneous graphs,
we introduce a new clustering-based anomaly detection approach called \method~ that 
($i$) can handle temporal graphs with typed nodes and edges,  
($ii$) processes incoming edges fast and consumes bounded memory, as well as 
($iii$) dynamically maintains the clustering and detects anomalies in real time.
In a nutshell, we propose a new shingling-based similarity function for heterogeneous graphs,
which lends itself to graph sketching that uses fixed memory while preserving similarity.
We show how to maintain the graph sketches efficiently as new edges arrive.
Based on this representation,
we employ and dynamically maintain a centroid-based clustering scheme to score and 
flag anomalous graphs.
The main contributions of this work are listed as follows:

\vspace{-0.065in}
\bit
\setlength{\itemsep}{-0.5\itemsep}

\item {\bf Novel formulation and graph similarity:}
We formulated the host-level APT detection problem 
as a clustering-based anomaly detection task in
streaming heterogeneous graphs.
To enable an effective clustering, 
we designed a new similarity function for timestamped
typed graphs, based on \emph{shingling}, which accounts for the frequency of
different substructures in a graph.
Besides being efficient to compute and effective in capturing similarity between graphs, the proposed function 
lends itself to comparing two graphs based on their {\em sketches}, which enables 
memory-efficiency.

\item {\bf Dynamic maintenance:} 
We introduce efficient techniques to keep the various components of our approach 
up to date as new edges arrive over the stream.
Specifically, we show how to maintain ($a$) the graph sketches, and
$(b)$ the clustering incrementally.

\item {\bf Desirable properties:} Our formulation and proposed techniques are motivated by the
requirements and desired properties of the application domain. As such, our approach is
($i$) {\em fully streaming}, where we perform a continuous, edge-level processing of the stream, rather than taking a snaphot-oriented approach; 
($ii$) {\em time-efficient}, where the processing of each edge is fast with constant complexity to update its graph's sketch and the clustering; 
($iii$) {\em memory-efficient}, where the sketches and cluster summaries consume constant memory that is controlled by user input, and
($iv$) {\em online}, where we score and flag the anomalies in real-time.

\eit
\vspace{-0.05in}

We quantitatively validate the effectiveness and (time and space) efficiency of our proposed \method~on simulated
datasets containing  normal host-level activity as well as abnormal attack scenarios (i.e., ground truth).
We also design experiments to study the approximation quality of our sketches, 
and the behavior of our detection techniques under varying parameters, such as memory size.

Source code of \method~ and the simulated
datasets (normal and attack) will be released at \url{http://www3.cs.stonybrook.edu/~emanzoor/streamspot/}.

\section{Problem \& Overview}
\label{sec:overview}

\begin{figure}[!t]
	\centering
	\begin{tabular}{c}
		\hspace{-0.25in}	\includegraphics[width=3.55in, height=0.9in]{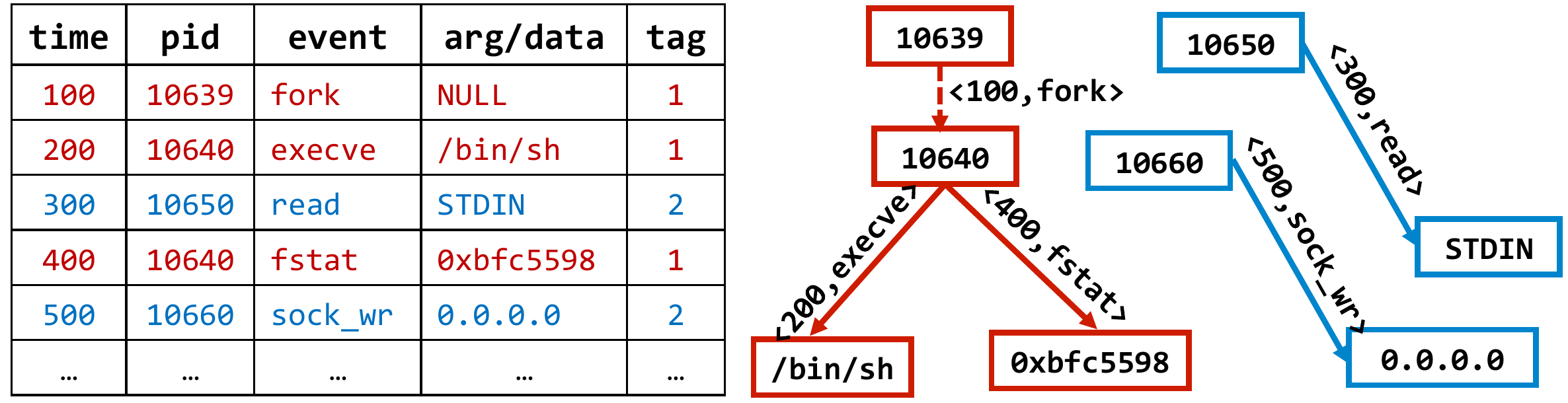}
	\end{tabular}
	\vspace{-0.125in}
	\caption{Example stream of system logs, and two resulting information flow graphs (red vs. blue).
		Both nodes and edges are typed. Edges arriving to different flows may interleave. \label{fig:flow}}
	\vspace{-0.15in}
\end{figure}

For host-level APT detection,
a host machine is instrumented to collect system
logs.
These logs essentially capture the events occuring in the system, such as memory accesses, system calls, etc.
An example log sequence is illustrated in Figure \ref{fig:flow}.
Based on the control and data dependences,
information flow graphs are constructed from the system logs.
In the figure, the {\tt tag} column depicts the ID of the information flow (graph) that an event (i.e., edge) belongs to.

The streaming graphs are heterogeneous where edge types correspond to system calls such as
{\tt read}, {\tt fork}, {\tt sock\_wr}, etc. and node types include
{\tt socket}, {\tt file}, {\tt memory}, etc.

As such, an edge can be represented in the form of

\vspace{0.05in}
{\tt< source-id, source-type $\phi_s$, dest-id, dest-type $\phi_d$, timestamp $t$, edge-type $\phi_e$,  flow-tag >}
\vspace{0.05in}

\sloppy{
	These edges form dynamically evolving graphs,
	where the edges sharing the same flow-tag belong to the same graph.
	Edges arriving to different graphs may be interleaved, that is, multiple graphs may be evolving simultaneously.
}

Our goal is to detect anomalous graphs at any given time $t$, i.e., in real time as they occur.
To achieve this goal, we follow a clustering-based anomaly detection approach.
In a nutshell, our method maintains a small, memory-efficient representation of the evolving graphs in main memory, and uses a new similarity measure that we introduce to cluster the graphs.
The clustering is also maintained dynamically as existing graphs evolve or as new ones arrive.
Anomalies are flagged in real time through deviations from this clustering model that captures the normal flow patterns.

In the following, we summarize the main components of our proposed approach called \method, with forward references to respective subsequent (sub)sections.

\vspace{-0.075in}
\bit
\setlength{\itemsep}{-0.5\itemsep}
\item {\bf Similarity of heterogeneous graphs:} (\S\ref{ssec:sim})
We introduce a new similarity measure for heterogeneous graphs with typed nodes and edges as well as timestamped edges. Each graph $G$ is represented by a what we call shingle-frequency vector (or shortly {\em shingle vector}) $\z_G$. Roughly, a $k$-shingle $s(v,k)$ is a string constructed by traversing edges, in their temporal order, in the $k$-hop neighborhood of node $v$. The shingle-vector contains the counts of unique shingles in a graph.
Similarity between two graphs is defined as the cosine similarity between their respective shingle  vectors.
Intuitively, the more the same shingles two graphs contain in common, the more similar they are.

\item {\bf Memory-efficient sketches:} (\S\ref{ssec:mes})
Number of unique shingles can be arbitrarily large for heterogeneous graphs with hundreds to thousands of node and edge types.
As such, we show how to instead use a {\em sketch} representation of a graph.
Sketches are much smaller (in fact constant-size) vectors, while enabling similarity to be preserved.
In other words, similarity of the sketches of two graphs provides a good approximation to their (cosine) similarity with respect to their original shingle vectors.

\item {\bf Efficient maintenance of sketches:} (\S\ref{ssec:ems})
As new edges arrive, shingle counts of a graph change.
As such, the shingle vector entries need to be updated.
Recall that we do not explicitly maintain this vector in memory, but rather its (much smaller) sketch.
In this paper, we show how to update the sketch of a graph efficiently, ($i$) in \emph{constant} time and ($ii$) without incurring \emph{any} additional memory overhead.

\item {\bf Clustering graphs (dynamically):} (\S\ref{sec:clustering})
We employ 
a centroid-based clustering of the graphs to capture normal behavior.
We show how to update the clustering as the graphs change
and/or
as new ones emerge, that again, exhibit small memory footprints.

\item {\bf Anomaly detection (in real time):} 
We score an incoming or updated graph by  its distance to the closest centroid
in the clustering.
Based on a distribution of distances for the corresponding cluster,
we quantify the significance of the score to flag anomalies.
Distance computation is based on (fixed size) sketches, as
such, scoring is fast for real time detection.
\eit

\hide{
	\begin{table}[t]
		\vspace{-0.1in}
		\caption{Frequently used symbols.}
		\centering
		\begin{tabular}{ll}
			\toprule
			\textbf{Symbol} & \textbf{Definition} \\
			\midrule
			$G$                     & a time-evolving graph\\
			$V, E$                  & sets of nodes and edges of a graph\\
			$k$                     & number of hops used in shingle construction\\
			$S$                     & the universe of unique shingles\\
			$L$                     & number of {\simhash} random vectors\\
			& size of the {\streamhash}/{\simhash} sketch\\
			& size of the projection vector\\
			$h_1, \dots, h_L$  & {\streamhash} hash functions\\
			
			$\mathbf{x}_G$          & sketch (vector) of graph $G$\\
			$\mathbf{y}_G$          & projection vector of graph $G$\\
			$\mathbf{z}_G$          & shingle (frequency) vector of graph $G$\\
			
			$|s|_{\textrm{max}}$    & maximum possible length of a shingle\\
			$N$                     & maximum no. of edges kept in memory\\
			
			$K$											& no. of bootstrap clusters\\

			$\oplus$                & string concatenation operator\\
			$~\cdot$                & vector or hash-function dot-product operator\\
			\bottomrule
		\end{tabular}
		\label{tbl:notation}
		\vspace{-0.15in}
	\end{table}
}

\vspace{-0.15in}
\section{Sketching Typed Graphs}
\label{sec:sketch}
We require a method to represent and compute the similarity between heterogeneous graphs that also captures the temporal order of edges. The graph representation must permit efficient online updates and consume bounded space with new edges arriving in an infinite stream.

Though there has been much work on computing graph similarity efficiently, existing methods fall short of our requirements. Methods that require knowing node correspondence \cite{journals/jisa/PapadimitriouDG10,conf/sdm/FaloutsosKV13} are inapplicable, as are {graph kernels} that precompute a fixed space of substructures \cite{menchetti2005weighted, shervashidze2011weisfeiler, feragen2013scalable} to represent graphs, which is infeasible in a streaming scenario. Methods that rely on \emph{global} graph metrics \cite{conf/asonam/Berlingerio13} cannot accommodate edge order and are also inapplicable. {Graph-edit-distance}-based \cite{bunke1983inexact} methods approximate hard computational steps with heuristics that provide no error guarantees, and are hence unsuitable.

We next present a similarity function for heterogenous graphs that captures the local structure and respects temporal order of edges. The underlying graph representation permits efficient updates as new edges arrive in the stream and consumes bounded space, without needing to compute or store the full space of possible graph substructures.

\subsection{Graph Similarity by Shingling}
\label{ssec:sim}

Analogous to \emph{shingling} text documents into $k$-grams \cite{broder1997resemblance} to construct their vector representations, we decompose each graph into a set of \emph{$k$-shingles} and construct a vector of their frequencies. The similarity between two graphs is then defined as the cosine similarity between their $k$-shingle frequency vectors. We formalize these notions below.

\vspace{-0.05in}
\begin{definition}[\emph{$k$-shingle}]
	Given a graph $G = (V,E)$ and node $v \in V$, the $k$-shingle $s(v,k)$ is a string constructed via a $k$-hop breadth-first traversal starting from $v$ as follows:
	
	\vspace{-0.05in}
	\begin{enumerate}
		\setlength{\itemsep}{-1.0\itemsep}
		\item Initialize $k$-shingle as type of node $v$: $s(v,k) = \phi_v$.
		\item Traverse the outgoing edges from each node in the order of their timestamps, $t$.
		\item For each traversed edge $e$ having destination $w$, concatenate the types of the edge and the destination node with the $k$-shingle: $s(v,k) = s(v,k) \oplus \phi_e \oplus \phi_w$.
	\end{enumerate}
	\label{def:kshingle}
\end{definition}
\vspace{-0.1in}

We abbreviate the \textbf{O}rdered $\mathbf{k}$-hop \textbf{B}readth \textbf{F}irst \textbf{T}raversal performed during the $k$-shingle construction defined above as {\okbfs}. It is important to note that the $k$-hop neighborhood constructed by an {\okbfs} is \emph{directed}.


\begin{definition}[\emph{shingle (frequency) vector} $\z_G$]
	Given the $k$-shingle universe $S$ and a graph $G = (V,E)$, let $S_G = \{s(v,k),~\forall v \in V\}$ be the set of $k$-shingles of $G$. $\z_G$ is a vector of size $|S|$ wherein each element $\z_{G}(i)$ is the frequency of shingle $s_i \in S$ in $S_G$.
\end{definition}

\begin{figure}[h]
	\vspace{-0.05in}
	\centering
	\begin{tabular}{c}
		\includegraphics[width=2.75in, height=1.4in]{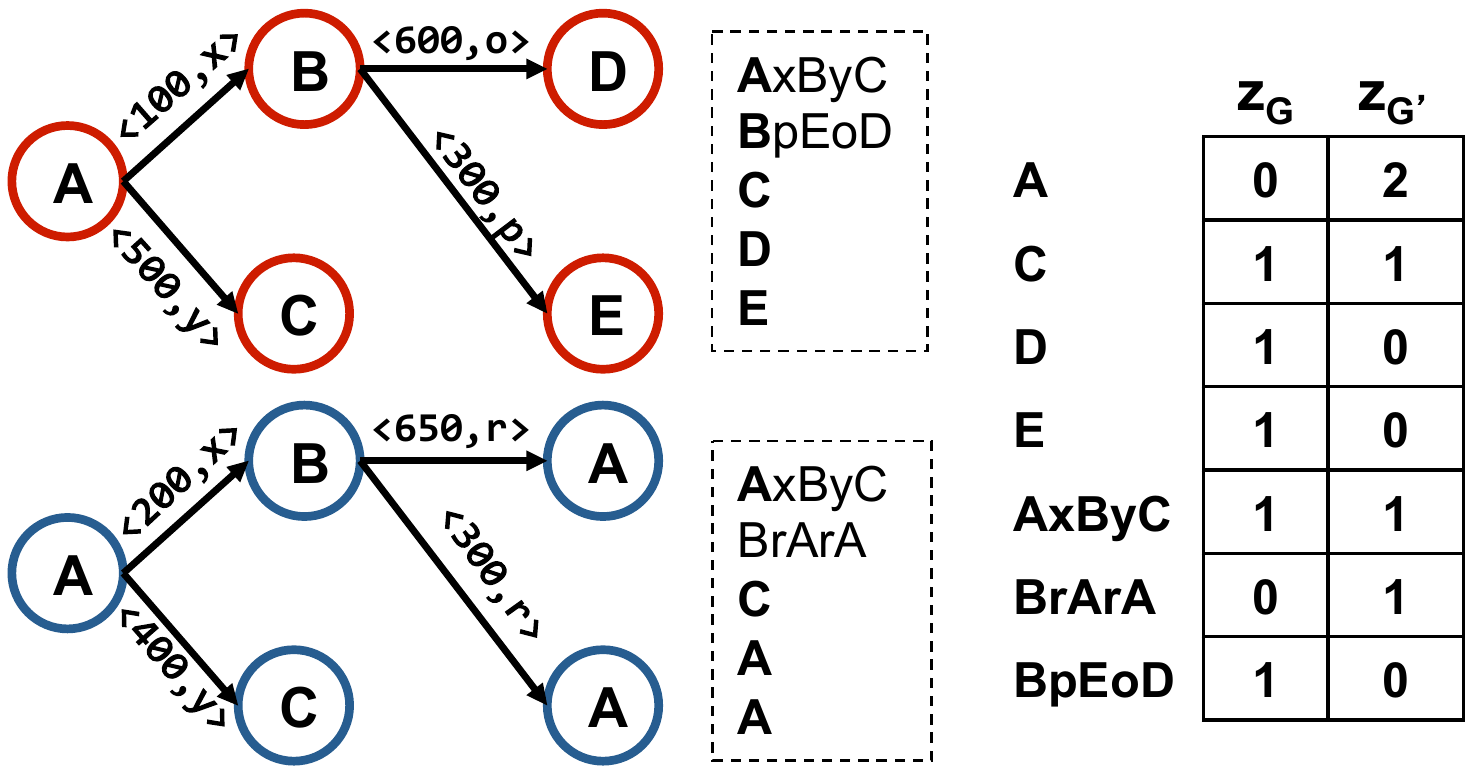}
	\end{tabular}
	\vspace{-0.1in}
	\caption{Shingling: (left) two example graphs with their shingles listed in dashed boxes (for $k=1$),
		(right) corresponding shingle frequency vectors.
		\label{fig:shingles}}
	\vspace{-0.025in}
\end{figure}

Shingling is illustrated for two example graphs in Figure \ref{fig:shingles}, along with their corresponding shingle vectors.

The similarity between two graphs $G$ and $G'$ is then the cosine similarity between their shingle vectors $\z_G$ and $\z_{G'}$.

Representing graphs by shingle vectors captures both their local graph structure and edge order. It also permits efficient online updates, since the local nature of $k$-shingles ensures that only a few shingle vector entries need to be updated for each newcoming edge (\S\ref{ssec:ems}). The parameter $k$ controls the trade-off between expressiveness and computational efficiency. A larger $k$ produces more expressive local neighborhoods, whereas a smaller one requires fewer entries to be updated in the shingle vector per incoming edge.


\subsection{Graph Sketches by Hashing}
\label{ssec:mes}

With a potentially large number of node and edge types, the universe $S$ of $k$-shingles may explode combinatorially and render it infeasible to store $|S|$-dimensional shingle vectors for each graph. We now present an alternate \emph{constant}-space graph representation that approximates the shingle count vector via \emph{locality-sensitive hashing} (LSH) \cite{indyk1998approximate}.

An LSH scheme for a given similarity function enables efficient similarity computation by projecting high-dimensional vectors to a low-dimensional space while preserving their  similarity. Examples of such schemes are {\minhash} \cite{broder1997resemblance} for the Jaccard similarity between sets and {\simhash} \cite{charikar2002similarity} for the cosine similarity between real-valued vectors, which we detail further in this section.

\subsubsection{{\simhash}}

Given input vectors in $\mathbb{R}^d$, {\simhash} is first instantiated with $L$ projection vectors $\mathbf{r}_1, \dots, \mathbf{r}_L \in \mathbb{R}^d$ chosen uniformly at random from the $d$-dimensional Gaussian distribution. The LSH $h_{\mathbf{r}_l}(\mathbf{z})$ of an input vector $\mathbf{z}$ for a given random projection vector $\mathbf{r}_l$, $l=1\ldots,L$, is defined as follows:
\begin{align}
	h_{\mathbf{r}_l}(\mathbf{z}) = \begin{cases} +1, \textrm{~if~} \mathbf{z}\cdot\mathbf{r}_l \ge 0\\
		-1, \textrm{~if~} \mathbf{z}\cdot\mathbf{r}_l < 0
	\end{cases}
	\label{eq:lsh-vector}
\end{align}

In other words $h_{\mathbf{r}_l}(\mathbf{z}) = sign(\mathbf{z}\cdot\mathbf{r}_l)$, which obeys the property that the probability (over vectors $\mathbf{r}_1, \dots, \mathbf{r}_L$) that any two input vectors $\mathbf{z}_G$ and $\mathbf{z}_{G'}$ hash to the same value is proportional to their cosine similarity:
\begin{align}
	\label{eq:prob}
	P_{l=1\ldots L}[h_{\mathbf{r}_l}(\mathbf{z}_G) = h_{\mathbf{r}_l}\big(\mathbf{z}_{G'}\big)] &\;=\; 1 - \frac{cos^{-1}(\frac{\mathbf{z}_G \cdot \mathbf{z}_{G'}}{\|\mathbf{z}_G\|\|\mathbf{z}_{G'}\|})}{\pi}
\end{align}

Since computing similarity requires only these hash values, each $d$-dimensional input vector $\mathbf{z}$ can be replaced with an $L$-dimensional \emph{sketch vector} $\x$ containing its LSH values, i.e., $\x = [h_{\mathbf{r}_1}(\mathbf{z}), \dots, h_{\mathbf{r}_L}(\mathbf{z})]$. As such, each sketch vector can be represented with just $L$ bits, where each bit corresponds to a value in $\{+1,-1\}$.

The similarity between two input vectors then can be estimated by empirically evaluating the probability in
Eq. \eqref{eq:prob} as the proportion of hash values that the input vectors agree on when hashed with $L$ random vectors.
That is,
\begin{align}
	sim(G,G')  & \propto \frac{|\{l: \x_G(l) = \x_{G'}(l)\}|}{L}
\end{align}

In summary, given a target dimensionality $L \ll |S|$, we can represent each graph $G$ with a sketch vector of dimension $L$, discard the $|S|$-dimensional shingle vectors and compute similarity in this new vector space.

\subsubsection{Simplified {\simhash}}

Note that we need to perform the \emph{same} set of random projections on the changing or newly emerging shingle vectors, with the arrival of new edges and/or new graphs.
As such, we would need to
maintain the set of $L$ projection vectors in main memory (for now, later we will show that we do not need them explicitly).

In practice,  the random projection vectors $\mathbf{r}_l$'s remain sufficiently random when each of their $|S|$ elements are drawn uniformly from $\{+1, -1\}$ instead of from a $|S|$-dimensional Gaussian distribution \cite{rajaraman2012mining}. With this simplification, just like the sketches, each projection vector can also be represented using $|S|$ \emph{bits}, rather than $32*|S|$ bits (assuming 4 bytes per float), further saving space.

\begin{figure}[!t]
	\centering
	\begin{tabular}{c}
		\includegraphics[width=3.25in, height=1.3in]{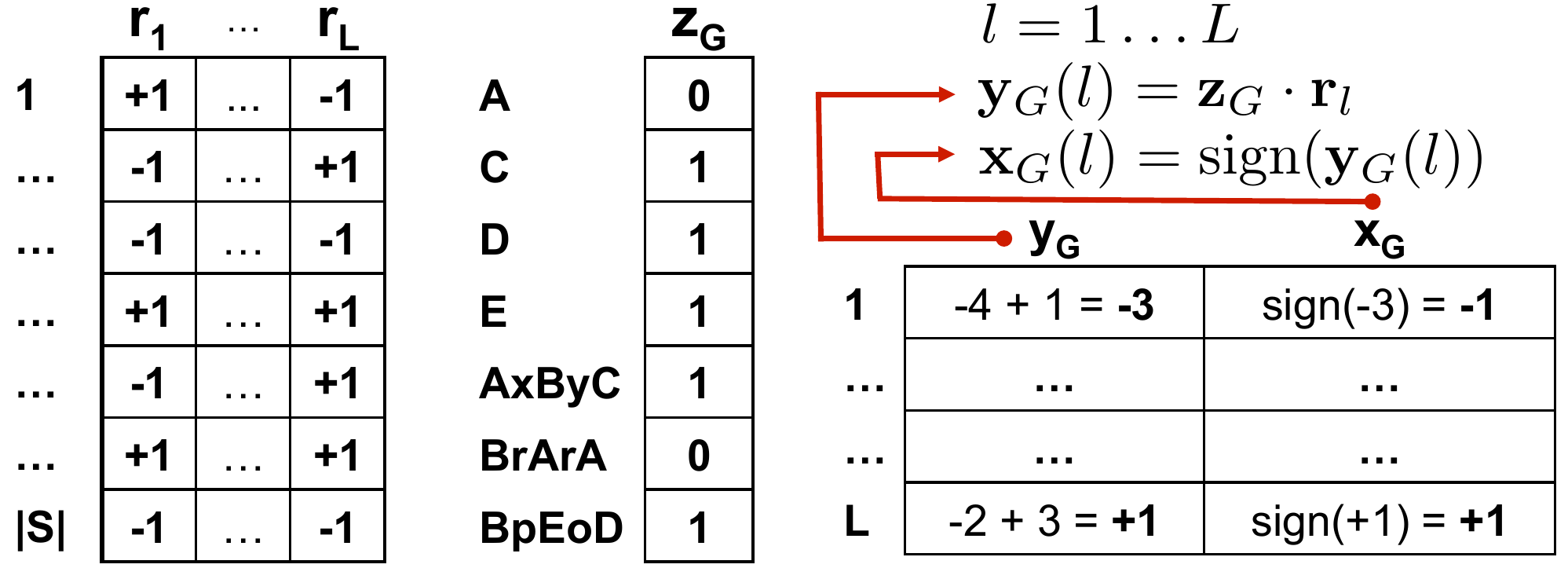}
	\end{tabular}
	\vspace{-0.1in}
	\caption{Sketching: (left) $L$ random vectors, (center) shingle vector $\z_G$ of graph $G$,
		(right) corresponding projection vector $\y_G$ and sketch vector $\x_G$.
		\label{fig:sketching}}
	\vspace{-0.15in}
\end{figure}

Figure \ref{fig:sketching} illustrates the idea behind sketching (for now in the static case).
Given $|S|$-dimensional random $\mathbf{r}_l$ vectors, $l=1\ldots L$, with elements in $\{+1,-1\}$ (left) and a shingle vector $\z_G$ (center), the $L$-dimensional sketch $\x_G$ is obtained by taking the sign of the dot product of $\z$ with each $\mathbf{r}_l$ (right).

However, three main shortcomings of {\simhash} remain: (1) it requires explicit projection vectors in memory,
(2) $|S|$ can still get prohibitively large, and (3) it requires knowing the size of the complete shingle universe $|S|$ to specify the dimension of each random vector. With new shingles continuously being formed from the new node and edge types arriving in the stream, the complete shingle universe (and hence its size) always remains \emph{unknown}. As such, {\simhash} as proposed cannot be applied in a streaming setting.

\subsubsection{{\streamhash}}
\label{sss:streamhash}

To resolve the issues with {\simhash} for the streaming setting,
we propose {\streamhash} which,  rather than $L$ $|S|$-dimensional random bit vectors (with entries corresponding to $\{+1,-1\}$ as described above), is instead instantiated with $L$ hash functions $h_1, \dots, h_L$ picked uniformly at random from a family $\mathcal{H}$ of hash functions, mapping shingles to $\{+1,-1\}$. That is, a $h\in \mathcal{H}$ is a \emph{deterministic} function that maps a fixed/given shingle $s$ to either $+1$ or $-1$.

\vspace{0.025in}
\textbf{Properties of $\mathcal{H}$.~} We require a family that exhibits three key properties; uniformity w.r.t. both shingles and hash functions, and pairwise-independence, as described below.

First, it should be
equally probable for a given shingle to hash to $+1$ or $-1$ over all hash functions in the family:
\begin{align}
	\textrm{Pr}_{h \in \mathcal{H}}[h(s) = +1] = \textrm{Pr}_{h \in \mathcal{H}}[h(s) = -1],~\forall s \in S.
	\label{eq:uniform-family}
\end{align}

Second, to disallow trivially uniform families such as $\mathcal{H} = \{\forall s \in S: h_1(s) = +1,~h_2(s) = -1\}$, we also require that for a given hash function, hash values in $\{+1,-1\}$ are equiprobable over all shingles in the universe:
\begin{align}
	\textrm{Pr}_{s \in S}[h(s) = +1] = \textrm{Pr}_{s \in S}[h(s) = -1],~\forall h \in \mathcal{H}.
	\label{eq:uniform-function}
\end{align}

To further disallow uniform families with correlated uniform hash functions such as $\mathcal{H} = \{h_1(s),~h_2(s) = -h_1(s)\}$ (where $h_1$ is some uniform hash function), we require the hash functions in the family to be {pairwise-independent}:
\begin{gather}
	\forall s,s' \in S \textrm{~s.t.~} s \ne s' \textrm{~and~} \forall t,t' \in \{+1,-1\},\nonumber\\
	\textrm{Pr}_{h \in \mathcal{H}}[h(s') = t' | h(s) = t] = \textrm{Pr}_{h \in \mathcal{H}}[h(s') = t'].
	\label{eq:independence}
\end{gather}

If the shingle universe is fixed and known, picking a hash function $h_l$ at random from $\mathcal{H}$ is equivalent to picking some vector $\mathbf{r}_l$ at random from $\{+1,-1\}^{|S|}$, with each element $\mathbf{r}_{l}(i)$ equal to the hash value $h_l(s_i)$.

If we overload the ``dot-product'' operator for an input vector $\mathbf{z}$ and a hash function $h_l$ as follows:
\begin{align}
	\mathbf{y}(l) = \mathbf{z} \cdot h_l = \sum_{i = 1, \dots, |S|} \mathbf{z}(i) h_l(s_i), \;\; l=1\ldots L
	\label{eq:hash-dot-product}
\end{align}
we can define the LSH $g_{h_l}(\mathbf{z})$ of the input vector $\z$ for the given hash function similar to Eq. \eqref{eq:lsh-vector}:
\begin{align}
	g_{h_l}(\mathbf{z}) = \begin{cases} +1, \textrm{~if~} \mathbf{z} \cdot h_l \ge 0\\
		-1, \textrm{~if~} \mathbf{z} \cdot h_l < 0
	\end{cases}
	\label{eq:lsh}
\end{align}

The $\mathbf{y}$ vector is called the \emph{projection vector} of a graph.
Each entry $\mathbf{y}(l)$ as given in Eq. \eqref{eq:hash-dot-product} essentially holds the sum of the counts of shingles that map to $+1$ by $h_l$ minus the sum of the counts of shingles that map to $-1$ by $h_l$.

The $L$-bit sketch can then be constructed for each input vector $\z$ by $\x=sign(\y)$ and used to compute similarity the same way as in {\simhash}. Unlike {\simhash}, the sketches in {\streamhash} can be constructed and maintained \emph{incrementally} (\S\ref{ssec:ems}), as a result of which, we no longer need to know the complete shingle universe $S$ or maintain $|S|$-dimensional random vectors in memory.

\vspace{0.025in}
\textbf{Choosing $\mathcal{H}$.~~}
A family satisfying the aforementioned three properties is said to be \emph{strongly universal} \cite{wegman1981stronglyuniversal}.

For our scenario, we adopt a fast implementation of the strongly universal multilinear family for strings \cite{lemire2014strongly}. In this family, an input string $s$ (i.e., shingle) is divided into $|s|$ components (i.e., ``characters'') as $s = c_1c_2 \dots c_{|s|}$. A hash function $h_l$ is constructed by first choosing $|s|$ random numbers $m_1^{(l)}, \dots, m_{|s|}^{(l)}$ and then hashing $s$ as follows:

\vspace{-0.1in}
\begin{align}
	h_l(s) = 2 * \bigg( (m_1^{(l)} + \sum_{i=2}^{|s|} m_i^{(l)} * int(c_i)) ~\text{mod}~ 2 \bigg) - 1.
	\label{eq:fast-multilinear-hash}
\end{align}
where $int(c_i)$ is the ASCII value of character $c_i$ and $h_l(s)\in \{+1,-1\}$.
Note that the hash value for a shingle $s$ of length $|s|$ can be computed in $\Theta(|s|)$ time.

We represent each hash function by $|s|_{\textrm{max}}$ random numbers, where
$|s|_{\textrm{max}}$ denotes the maximum possible length of a shingle.
These numbers are fixed per hash function $h_l$, as it is a deterministic function that hashes a fixed/given shingle to the same value each time.
In practice, $L$ hash functions can be chosen uniformly at random from this family by generating $L*|s|_{\textrm{max}}$ uniformly random 64-bit integers using a pseudorandom number generator.

\vspace{0.025in}
\textbf{Merging sketches.~~} Two graphs $G$ and $G'$ will merge if an edge arrives in the stream having its source node in $G$ and destination node in $G'$, resulting in a graph that is their union $G \cup G'$. The centroid of a cluster of graphs is also represented by a function of their union (\S\ref{sec:clustering}). Both scenarios require constructing the \emph{sketch of the union of graphs}, which we detail in this subsection.

The shingle vector of the union of two graphs $G$ and $G'$ is the sum of their individual shingle vectors:
\begin{align}
	\z_{G \cup G'} = \z_G + \z_{G'}.
	\label{eq:union-shingle-vector}
\end{align}

As we show below, the projection vector of the union of two graphs $\mathbf{y}_{G \cup G'}$ also turns out to be the sum of their individual projection vectors $\y_{G}$ and $\y_{G'}$; $\forall l = 1, \dots, L$:
\begin{align}
	\y_{G \cup G'}(l) &= \z_{G \cup G'} \cdot h_l
	\tag{by Eq. \eqref{eq:hash-dot-product}}\nonumber\\
	&= \sum_{i = 1, \dots, |S|} \left(\z_{G}(i) + \z_{G'}(i)\right) h_l(s_i)
	\tag{by Eq. \eqref{eq:union-shingle-vector}}\nonumber\\
	&= \sum_{i = 1, \dots, |S|} \z_{G}(i)h_l(s_i) + \sum_{i = 1, \dots, |S|}\z_{G'}(i) h_l(s_i)\nonumber\\
	&= \y_{G}(l) + \y_{G'}(l).
\end{align}

Hence, the $L$-bit sketch of $G \cup G'$ can be computed as $\x_{G \cup G'} = sign(\y_{G} + \y_{G'})$. This can trivially be extended to construct the sketch of the union of any number of graphs.


\subsection{Maintaining Sketches Incrementally}
\label{ssec:ems}

We now describe how {\streamhash} sketches are updated on the arrival of a new edge in the stream. Each new edge being appended to a graph gives rise to a number of \emph{outgoing} shingles, which are removed from the graph, and \emph{incoming} shingles, which are added to the graph. These shingles are constructed by {\okbfs} traversals from certain nodes of the graph, which we detail further in this section.

Let $e(u,v)$ be a new edge arriving in the stream from node $u$ to node $v$ in some graph $G$. Let $\mathbf{x}_G$ be the $L$-bit sketch vector of $G$. We also associate with each graph a length-$L$ \emph{projection vector} $\mathbf{y}_G$, which contains ``dot-products'' (Eq. \eqref{eq:hash-dot-product}) for the hash functions $h_1, \dots, h_L$. For an empty graph, $\y_G=\mathbf{0}$ and $\mathbf{x}_G = \mathbf{1}$ since $\z(i)$'s are all zero.

For a given incoming shingle $s_i$, the corresponding $\z(i)$ implicitly{\footnote{As we do not maintain the shingle vector $\z$'s explicitly.}} increases by $1$. This requires each element of the projection vector $\mathbf{y}_{G}(l)$ to be updated by simply adding the corresponding hash value $h_l(s_i) \in \{+1,-1\}$ due to the nature of the dot-product in Eq. \eqref{eq:hash-dot-product}. Updating $\mathbf{y}_G$ for an outgoing shingle proceeds similarly but by subtracting the hash values. For each element $\mathbf{y}_{G}(l)$ of the projection vector that is updated \emph{and} that changes sign, the corresponding bit of the sketch $\mathbf{x}_{G}(l)$ is updated using the new sign (Eq. \eqref{eq:lsh}). Updating the sketch for an incoming or outgoing shingle $s$ is formalized by the following update equations. $\forall l = 1, \dots, L\nonumber$:

$\;$
\vspace{-0.2in}
\begin{align}
	\mathbf{y}_{G}(l) &= \begin{cases} \mathbf{y}_{G}(l) + h_l(s), \textrm{~if} ~s~ \textrm{an incoming shingle}\\
		\mathbf{y}_{G}(l) - h_l(s), \textrm{~if} ~s~ \textrm{an outgoing shingle}\\
	\end{cases}  	\\
	\mathbf{x}_{G}(l) &= {\mathrm{sign}}(\mathbf{y}_{G}(l)).
\end{align}

Now that we can update the sketch (using the updated projection vector) of a graph for both incoming and outgoing shingles, without maintaining any shingle vector explicitly, we need to describe the construction of the incoming and outgoing shingles for a new edge $e$.

Appending $e$ to the graph updates the shingle for every node that can reach $e$'s destination node $v$ in at most $k$ hops, due to the nature of $k$-shingle construction by {\okbfs}. For each node $w$ to be updated, the incoming shingle is constructed by an {\okbfs} from $w$ that considers $e$ during traversal, and the outgoing shingle is constructed by an {\okbfs} from $w$ that ignores $e$ during traversal. In practice, both shingles can be constructed by a single modified-{\okbfs} from $w$ parameterized with the new edge.

Since the incoming shingle for a node may be the outgoing shingle for another, combining and further collapsing the incoming and outgoing shingles from all the updated nodes will enable updating the sketch while minimizing the number of redundant updates.

\subsection{Time and Space Complexity }
\label{ss:complexity}
\textbf{Time.} Since sketches are constructed incrementally (\S\ref{ssec:ems}), we evaluate the running time for each new edge arriving in the stream. This depends on the largest directed $k$-hop neighborhood possible for the nodes in our graphs. Since the maximum length of a shingle $|s|_{\textrm{max}}$ is proportional to the size of this neighborhood, we specify the time complexity in terms of $|s|_{\textrm{max}}$.

A new edge triggers an update to $O(|s|_{\textrm{max}})$ nodes, each of which results in an {\okbfs} that takes $O(|s|_{\textrm{max}})$ time. Thus, it takes $O(|s|^2_{\textrm{max}})$ time to {\tt construct} the $O(|s|_{\textrm{max}})$ incoming and outgoing shingles for a new edge. Hashing each shingle takes $O(|s|_{\textrm{max}})$ time (\S\ref{sss:streamhash}) resulting in a total {\tt hashing} time of $O(|s|^2 _{\textrm{max}})$. Updating the projection vector elements and bits in the sketch takes $O(L)$ time.

This leads to an overall sketch update time of $O(L + |s|^2 _{\textrm{max}})$ per edge. Since $L$ is a constant parameter and $|s|_{\textrm{max}}$ depends on the value of the parameter $k$, the per-edge running time can be controlled.

\vspace{0.005in}
\textbf{Space.} Each graph (of size at most $|G|_{\textrm{max}}$) with its sketch and projection vectors consumes $O(L + |G|_{\textrm{max}})$ space.{\footnote{Note that the projection vector holds $L$ positive and/or negative \emph{integers}, and the sketch is a length-$L$ \emph{bit} vector.}} However, the number of graphs in the stream is unbounded, as such the overall space complexity is dominated by storing graphs. Hence, we define a parameter $N$ to limit the maximum number of edges we retain in memory at any instant. Once the total number of edges in memory exceeds $N$, we evict the oldest edge incident on the least recently \emph{touched} node. The rationale is that nodes exhibit locality of reference by the edges in the stream that touch them (i.e., that have them as a source or destination).
With up to $N$ edges, we also assume a {\em constant} number $c$ of graphs is maintained and processed in memory at any given time.

The total space complexity is then $O(cL + N)$ which can also be controlled. Specifically, we choose $N$ proportional to the available memory size, and $L$ according to the required quality of approximation of graph similarity.


\section{Anomaly Detection}
\label{sec:detection}

\label{sec:clustering}

\textbf{Bootstrap Clusters.~~} \method ~is first initialized with bootstrap clusters obtained from a training dataset of benign flow-graphs. The training graphs are grouped into $K$ clusters using the $K$-medoids algorithm, with $K$ chosen to maximize the silhouette coefficient~\cite{Rousseeuw87} of the resulting clustering. This gives rise to compact clusters that are well-separated from each other. An \emph{anomaly threshold} for each cluster is set to 3 standard deviations greater than the mean distance between the cluster's graphs and medoid. This threshold is derived from Cantelli's inequality~\cite{grimmett2001probability} with an upper-bound of 10\% on the false positive rate.

Provided the bootstrap clusters, \method ~constructs \streamhash ~projection vectors for each training graph, and constructs the projection vector of the centroid of each cluster as the average of the projection vectors of the graphs it contains. In essence, the centroid of a cluster is the ``average graph'' with shingle vector counts formed by the union (\S\ref{sss:streamhash}) of the graphs it contains divided by the number of graphs. The sketch of each centroid is then constructed and the bootstrap graphs are discarded from memory.

\vspace{0.025in}
\textbf{Streaming Cluster Maintenance.~~} Apart from the cluster centroid sketches and projection vectors, we maintain in memory the number of graphs in each cluster and, for each observed and unevicted graph, its anomaly score and assignment to either one of $K$ clusters or an ``attack'' class. Each new edge arriving at a graph $G$ updates its sketch $\x_G$ and projection vector $\y_G$ to $\x'_G$ and $\y'_G$ respectively (\S\ref{ssec:ems}). $\x'_G$ is then used to compute the distance of $G$ to each cluster centroid. Let $Q$ be the nearest cluster to $G$, of size $|Q|$ and with centroid sketch $\x_Q$ and centroid projection vector $\y_Q$.

If $G$ was previously unassigned to any cluster and the distance of $G$ to $Q$ is lesser than its corresponding cluster threshold, then $G$ is assigned to $Q$ and its size and projection vector are updated $\forall l = 1, \dots, L$ as: 
\begin{align}
	\label{eq:cluster-add}
	\y_{Q}(l) & = \frac{\y_{Q}(l) \times |Q| + \y'_{G}(l)}{|Q| + 1}, \;\;\;\; \; |Q| = |Q| + 1 \;.
\end{align}
If the graph was previously already assigned to $Q$, its size remains the same and its projection vector is updated as:
\begin{align}
	\label{eq:cluster-update}
	\y_{Q}(l) &= \y_{Q}(l) + \frac{\y'_{G}(l) - \y_{G}(l)}{|Q|} \;.
\end{align}
If the graph was previously assigned to a different cluster $R \neq Q$, $Q$ is updated using Eq. \ref{eq:cluster-add} and the size and projection vector of $R$ are updated as:
\begin{align}
	\label{eq:cluster-remove}
	\y_{R}(l) & = \frac{\y_{R}(l) \times |R| - \y_{G}(l)}{|R| - 1},  \;\;\;\; |R| = |R| - 1 \;.
\end{align}
If the distance from $G$ to $Q$ is greater than its corresponding cluster threshold, $G$ is removed from its assigned cluster (if any) using Eq. \eqref{eq:cluster-remove} and assigned to the ``attack'' class. In all cases where the projection vector of $Q$ (or $R$) is updated, the corresponding sketch is also updated as:
\begin{align}
	\label{eq:cluster-sketch-update}
	\x_{Q}(l) &= {sign}(\y_{Q}(l)), \quad \forall l = 1, \dots, L.
\end{align}
Finally, the anomaly score of $G$ is computed as its distance to $Q$ after $Q$'s centroid has been updated.

\vspace{0.025in}
\textbf{Time and Space Complexity.~~} With $K$ clusters and $L$-bit sketches, finding the nearest cluster takes $O(KL)$ time and computing the graph's anomaly score takes $O(L)$ time. Adding a graph to (Eq. \eqref{eq:cluster-add}), removing a graph from (Eq. \eqref{eq:cluster-remove}) and updating (Eq. \eqref{eq:cluster-update}) a cluster each take $O(L)$ time, leading to a total time complexity of $O(KL)$ per-edge.

With a maximum of $c$ graphs retained in memory by limiting the maximum number of edges to $N$ (\S\ref{ss:complexity}), storing cluster assignments and anomaly scores each consume $O(c)$ space. The centroid sketches and projection vectors each consume $O(KL)$ space, leading to a total space complexity of $O(c + KL)$ for clustering and anomaly detection.

\section{Evaluation}
\label{sec:eval}

\textbf{Datasets.~~} Our datasets consist of flow-graphs derived from 1 attack and 5 benign scenarios. The benign scenarios involve normal browsing activity, specifically watching YouTube, downloading files, browsing \url{cnn.com}, checking Gmail, and playing a video game.  The attack involves a drive-by download triggered by visiting a malicious URL that exploits a Flash vulnerability and gains root access to the visiting host. For each scenario, Selenium Remote Control{\footnote{\url{www.seleniumhq.org/projects/remote-control/}}} was used to automate the execution of a 100 tasks. All system calls on the machine from the start of a task until its termination were traced and used to construct the flow-graph for that task. These flow-graphs were compiled into 3 datasets, the properties of which are shown in Table \ref{tbl:datasets}.

\begin{table*}
	\vspace{-0.1in}
	\caption{Dataset summary: Training scenarios and test edges (attack + 25\% benign graphs).}
	\centering
	\begin{tabular}{llcccc}
		\toprule
		\textbf{Dataset} & \textbf{Scenarios} & \textbf{\# Graphs} & \textbf{Avg. |V|} 			&  \textbf{Avg. |E|} 		& \textbf{\# Test Edges} \\
		\midrule
		\DA 							& YouTube, Download, CNN 								& 300 & 8705 & 239648 & 21,857,899 \\
		\DB 							& GMail, VGame, CNN 										& 300 & 8151 & 148414 & 13,854,229 \\
		\DC 							& YouTube, Download, CNN, GMail, VGame 	& 500 & 8315 & 173857 & 24,826,556 \\
		\bottomrule
	\end{tabular}
	\label{tbl:datasets}
	\vspace{-0.15in}
\end{table*}

\vspace{0.025in}
\textbf{Experiment Settings.~~}
We evaluate \method~in the following settings:

(1)  \emph{Static}:
We use $p$\% of all the benign graphs for training, and the rest of the benign graphs along with the attack graphs for testing. We find an offline clustering of the training graphs and then score and rank the test graphs based on this clustering. Graphs are represented by their shingle vectors and all required data is stored in memory. The goal is to quantify the effectiveness of \method~ before introducing approximations to optimize for time and space.



(2)  \emph{Streaming}:
We use $p$\% of the benign graphs for training to first construct a bootstrap clustering offline. This is provided to initialize \method, and the test graphs are then streamed in and processed online one edge at a time. Hence, test graphs may be seen only partially at any given time. For each edge, \method~updates the corresponding graph sketch, clusters, cluster assignments and anomaly scores, and a snapshot of the anomaly scores is retained every 10,000 edges for evaluation. \method~is also evaluated under memory constraints by limiting the sketch size and maximum number of stored edges.

%

\subsection{Static Evaluation}

We first cluster the training graphs based on their shingle-vector similarity. Due to the low diameter and large out-degree exhibited by flow-graphs, the shingles obtained tend to be long (even for $k=1$), and similar pairs of shingles from two graphs differ only by a few characters; this results in most pairs of graphs appearing dissimilar.

\begin{figure}[h!]
	\vspace{0.06in}
	\centering
		\begin{tabular}{ccc}
			\hspace{-0.1in}	{\includegraphics[width=.15\textwidth]{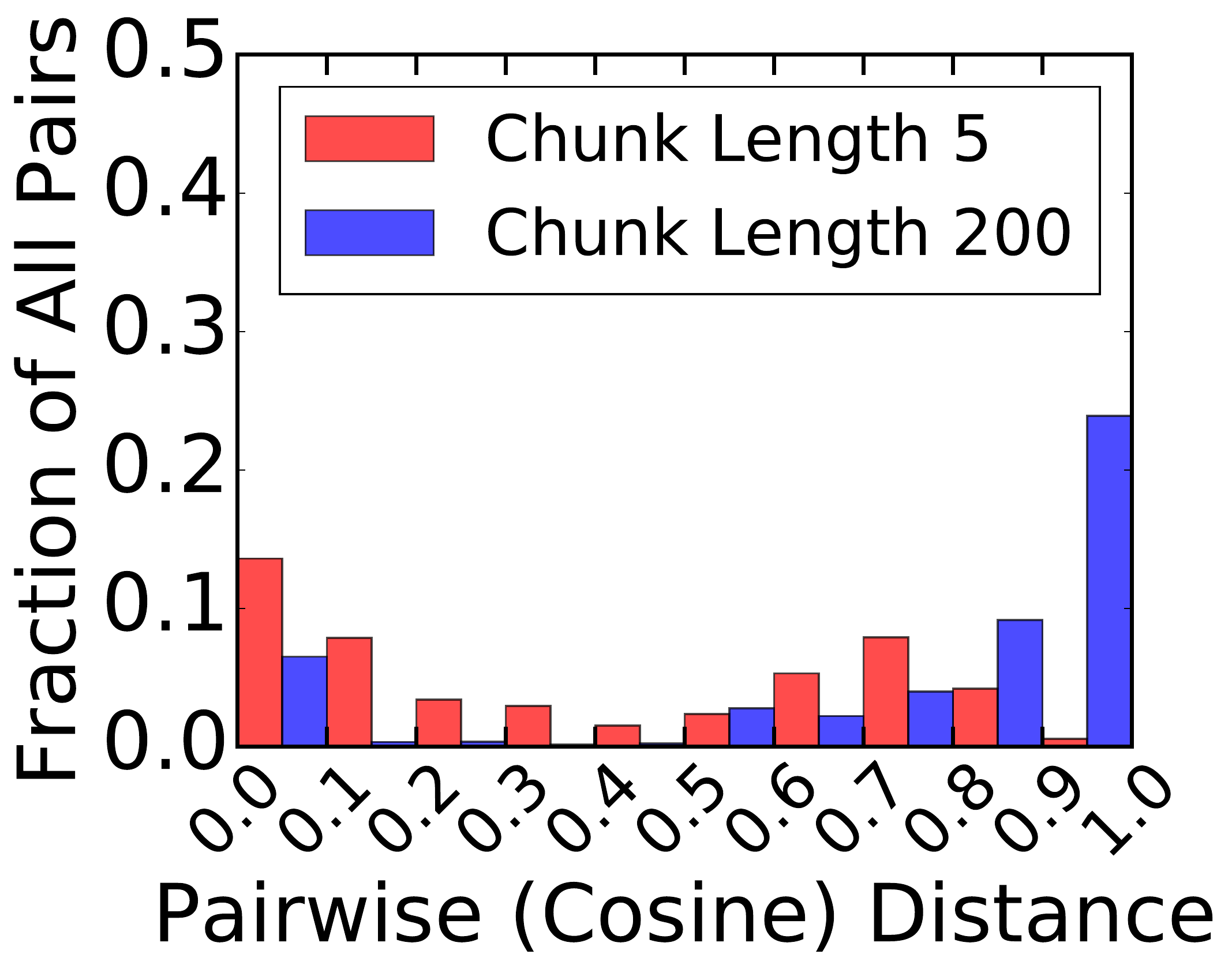}}&
			\hspace{-0.1in}	{\includegraphics[width=.15\textwidth]{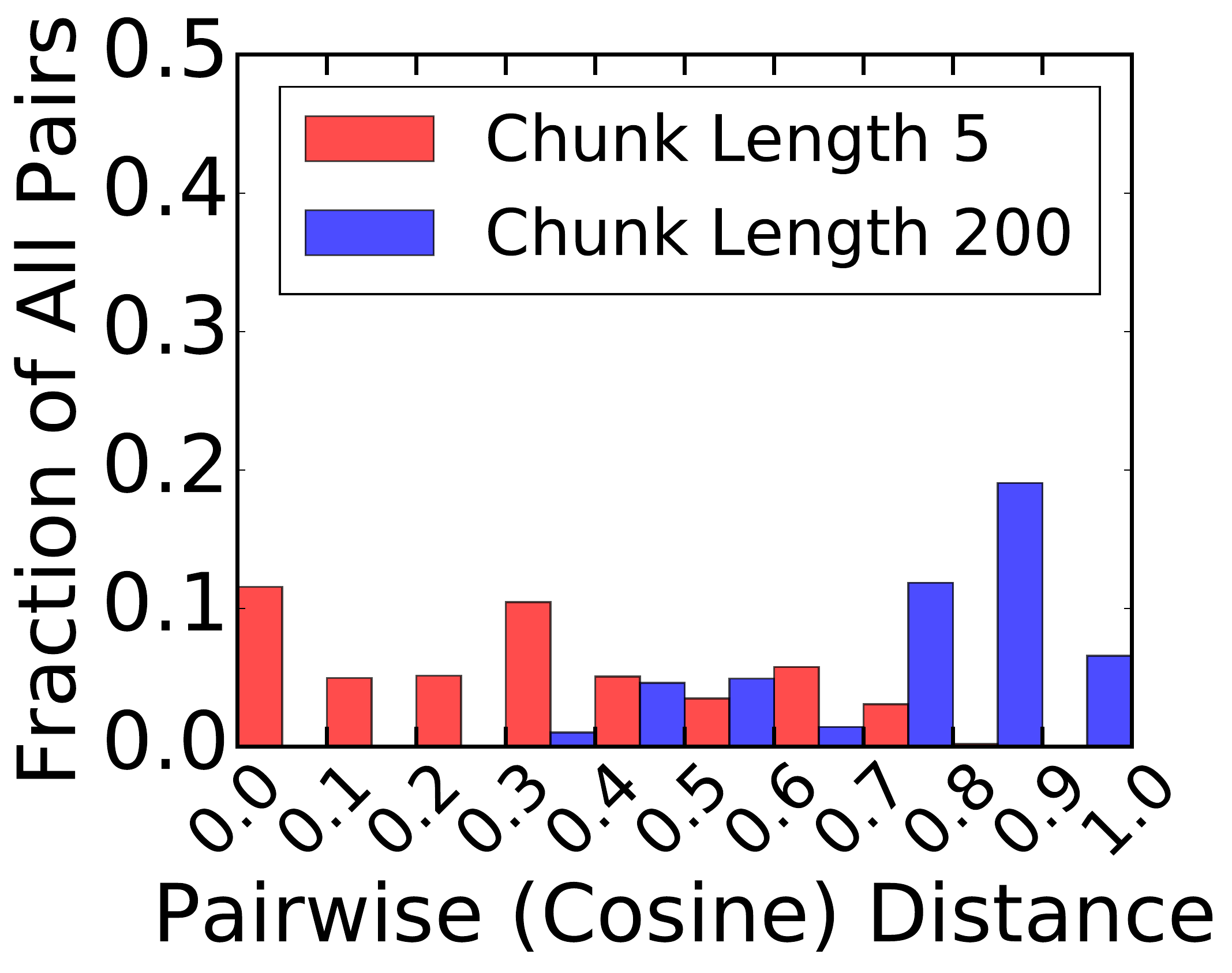}}&
			\hspace{-0.1in}	{\includegraphics[width=.15\textwidth]{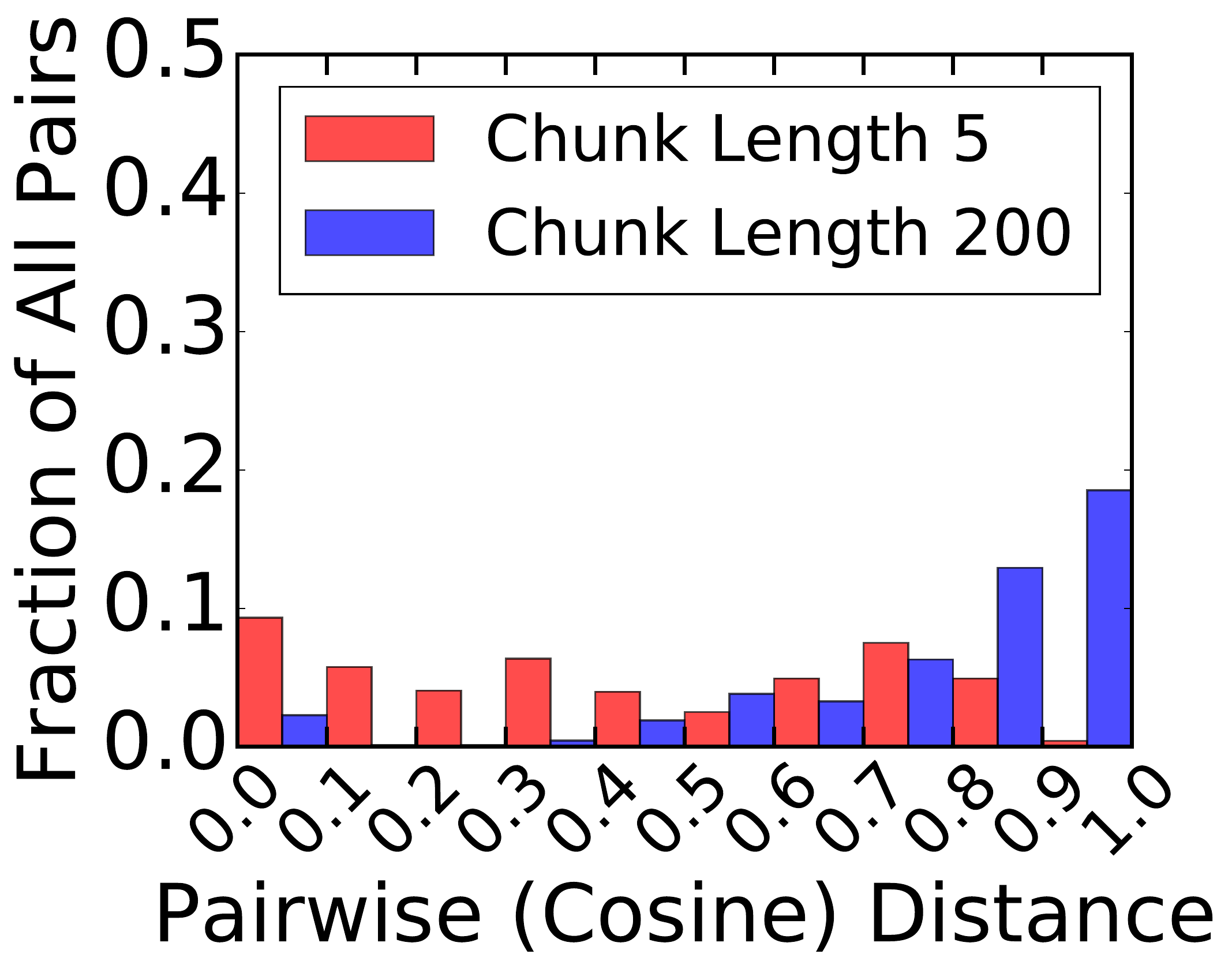}}\\
			(a) \DA & (b) \DB & (c) \DC 
		\end{tabular}
	\vspace{-5mm}
	\caption{Distribution of pairwise cosine distances for different values of chunk lengths.
		\label{fig:chunklen}}
	\vspace{-2.5mm}
\end{figure}

To mitigate this, we `chunk' each shingle by splitting it into fixed-size units. The chunk length parameter $C$ controls the influence of graph structure and node type frequency on the pairwise similarity of graphs. A small $C$ reduces the effect of structure and relies on the frequency of node types, making most pairs of graphs similar. A large $C$ tends to make pairs of graphs more dissimilar. This variation is evident in Figure \ref{fig:chunklen}, showing the pairwise-distance distributions for different chunk lengths.

\begin{figure}[h!]
	\centering
	\begin{tabular}{ccc}
\hspace{-0.1in}	{\includegraphics[width=.15\textwidth]{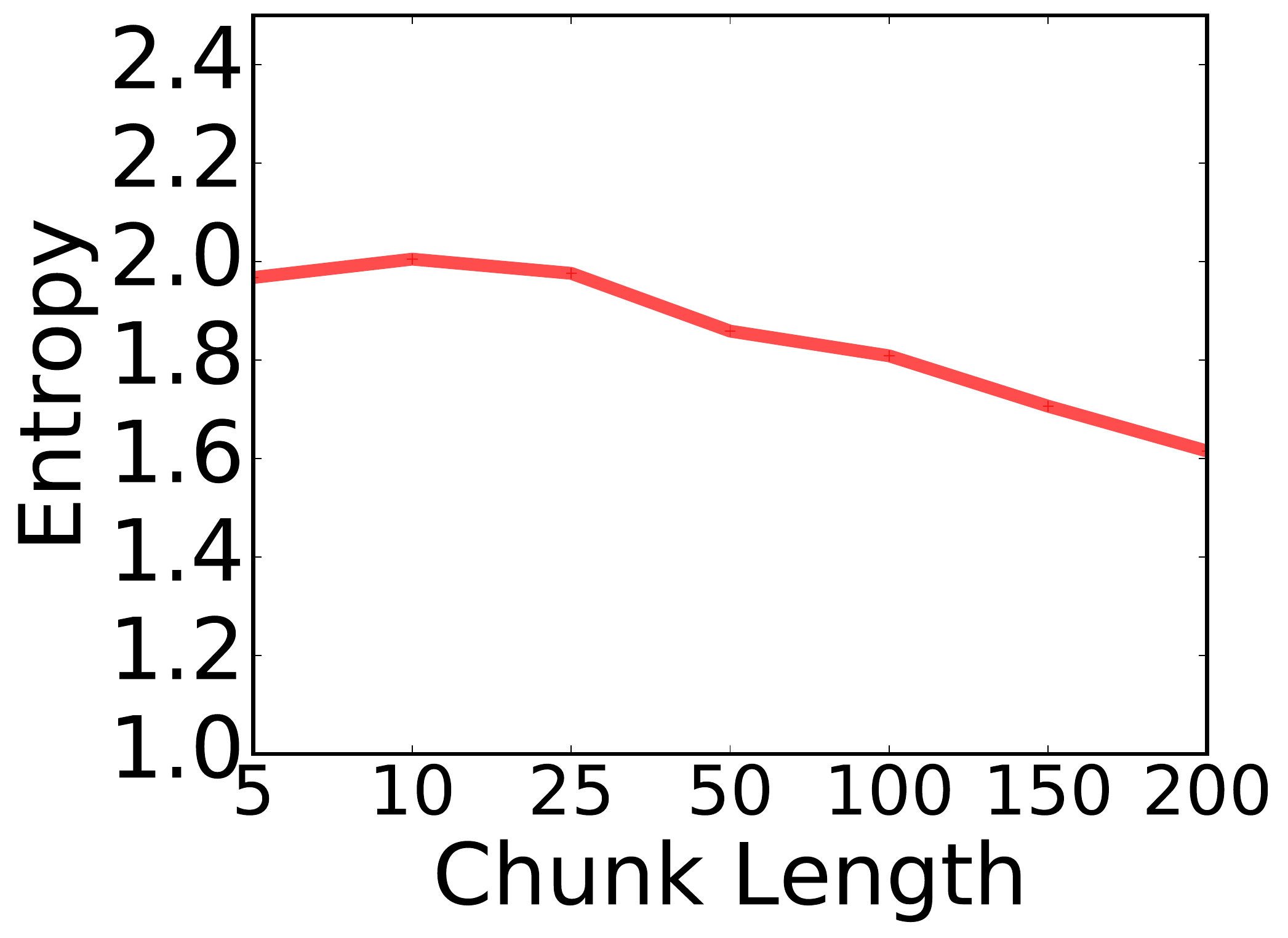}}&
\hspace{-0.1in}	{\includegraphics[width=.15\textwidth]{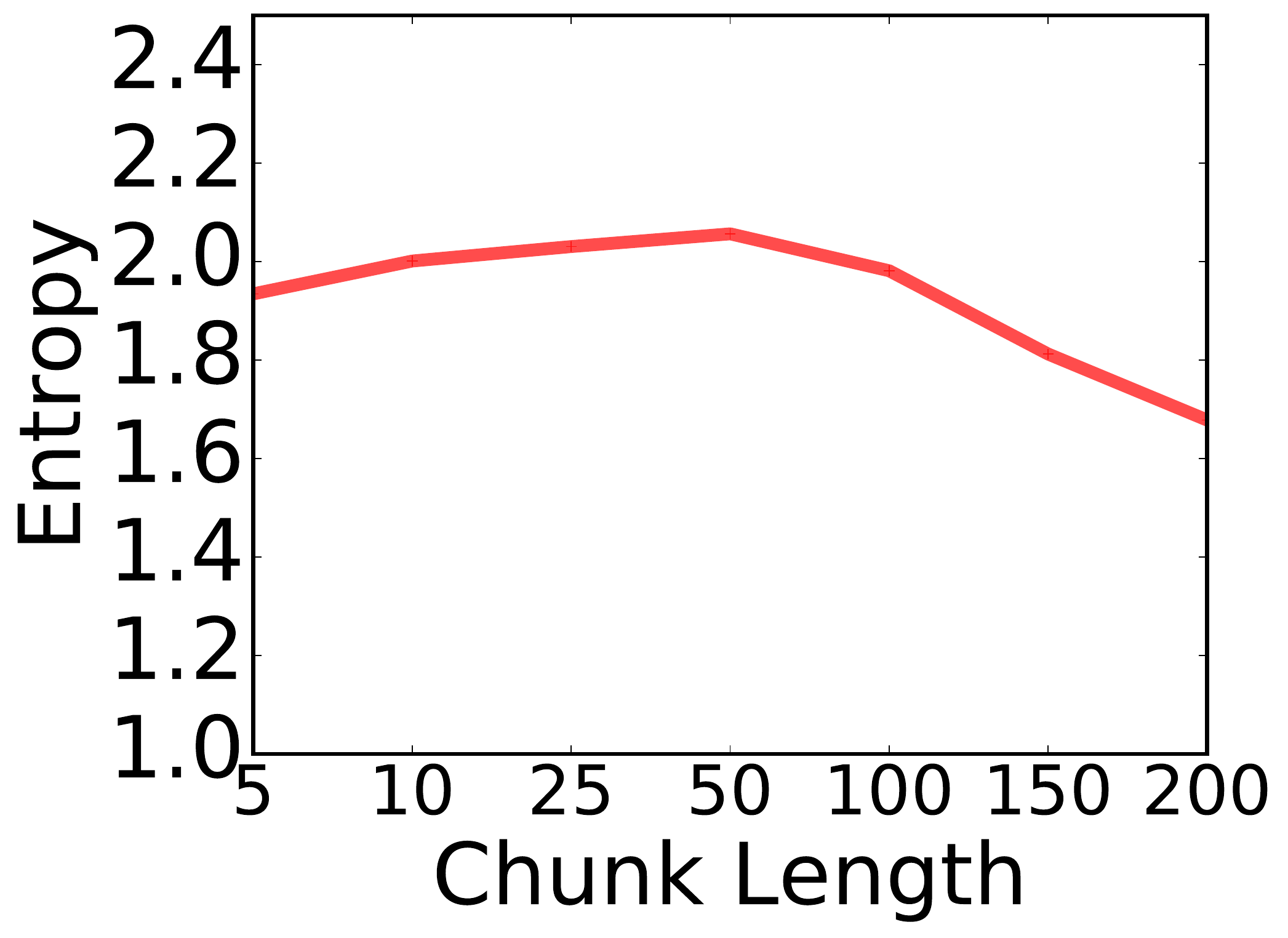}}&
\hspace{-0.1in}	{\includegraphics[width=.15\textwidth]{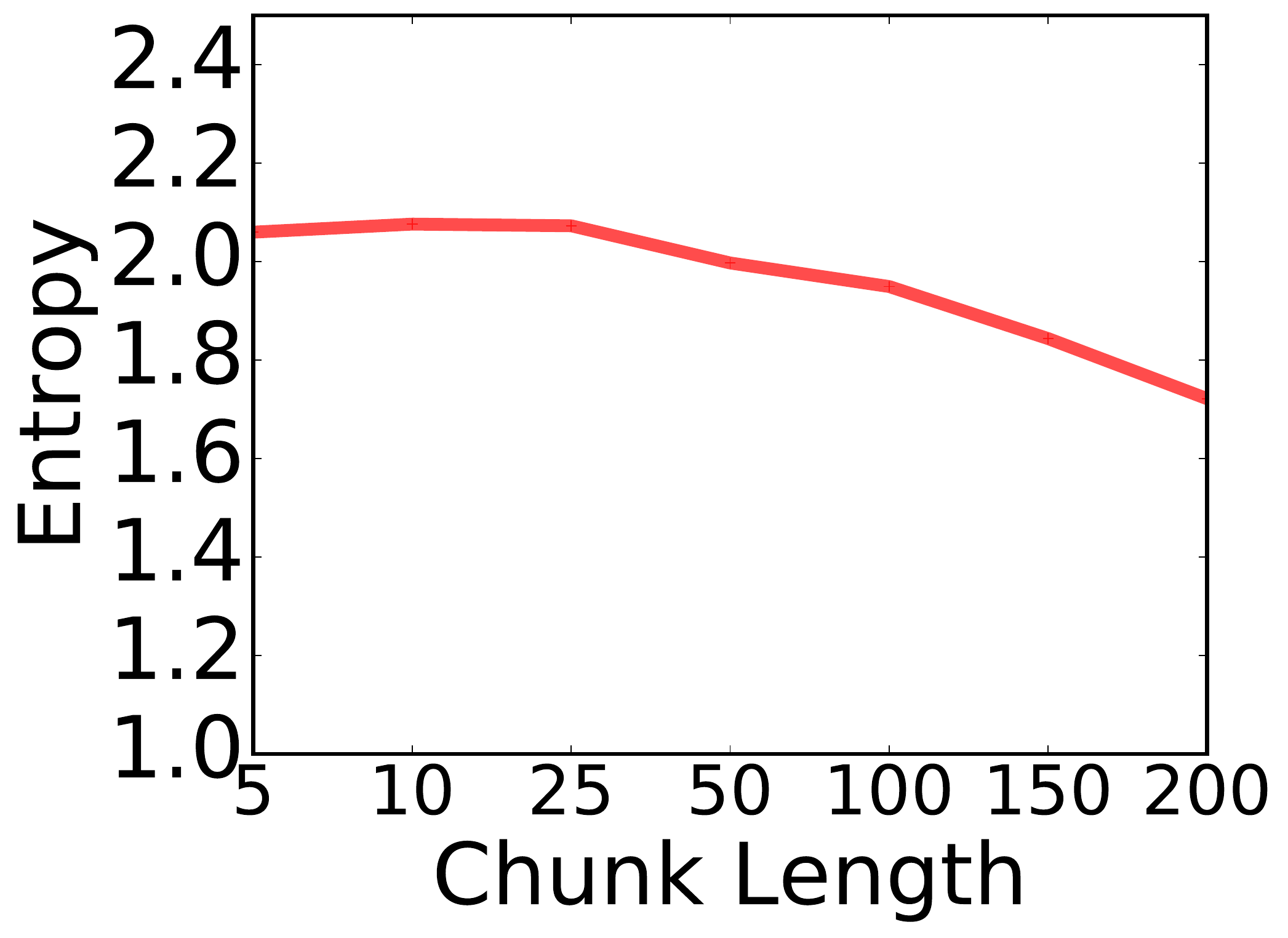}}\\
	(a) \DA & (b) \DB & (c) \DC 
	\end{tabular}
	\vspace{-5mm}
	\caption{Variation in entropy of the pairwise cosine distance distribution. ($p=75\%$)}
	\vspace{-1mm}
	\label{fig:pick-chunk-length}
\end{figure}

We aim to choose a $C$ that neither makes all pairs of graphs too similar or dissimilar. Figure \ref{fig:pick-chunk-length} shows the entropy of pairwise-distances with varying $C$ for each dataset. At the point of maximum entropy, the distances are near-uniformly distributed. A ``safe region'' to choose $C$ is near and to the right of this point; intuitively, this $C$ sufficiently differentiates dissimilar pairs of graphs, while not affecting similar ones. For our experiments, we pick $C=25, 100, 50$ respectively for \DA, \DB, and \DC. After fixing $C$, we cluster the training graphs with $K$-medoids and pick $K$ with the maximum silhouette coefficient for the resulting clustering; respectively $K=5, 5, 10$  for \DA, \DB, and \DC.

\begin{figure}[h!]
	\vspace{-0.1in}
	\centering
		\begin{tabular}{ccc}
			\hspace{-0.1in}	{\includegraphics[width=.15\textwidth]{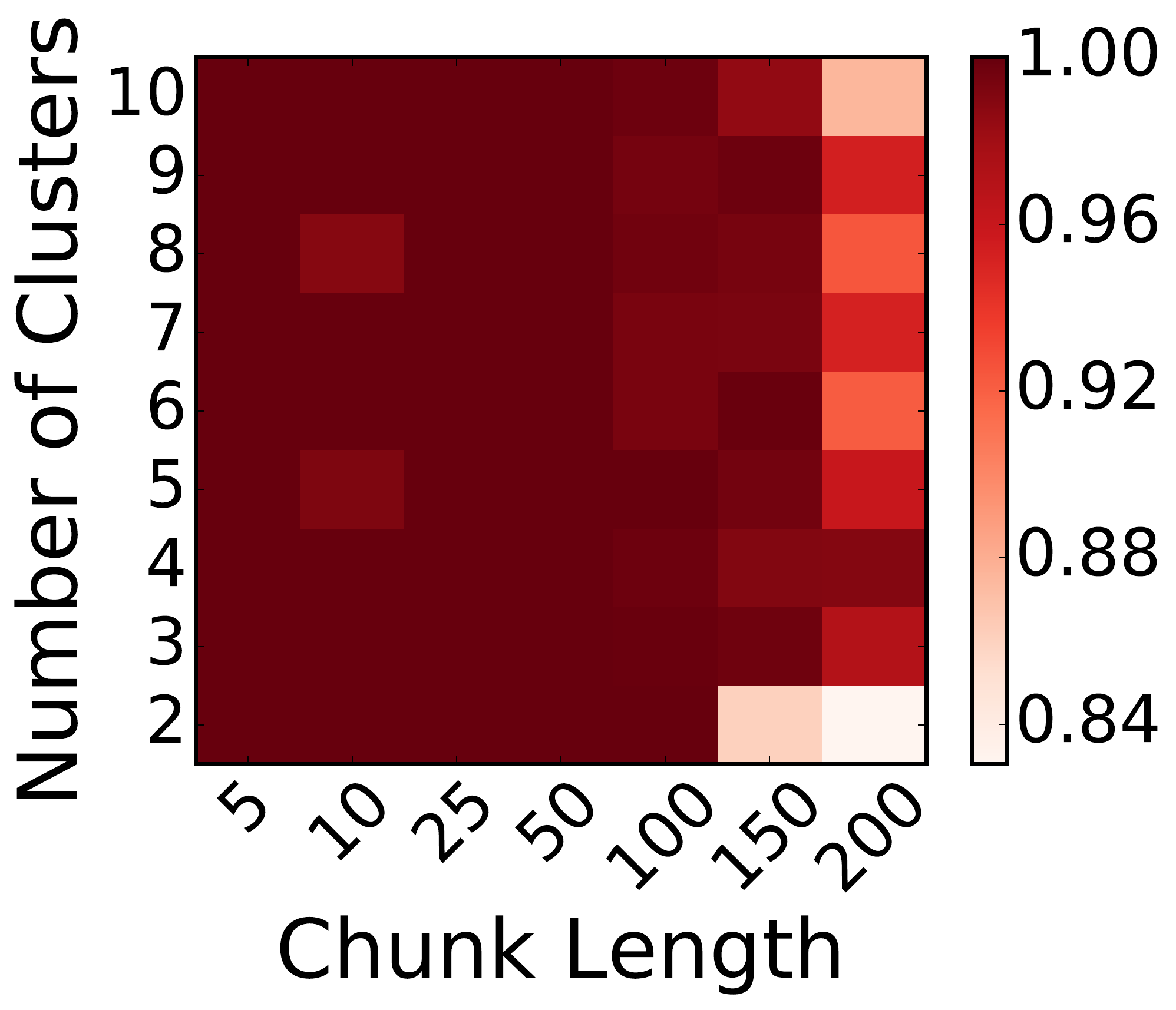}}&
			\hspace{-0.05in}	{\includegraphics[width=.15\textwidth]{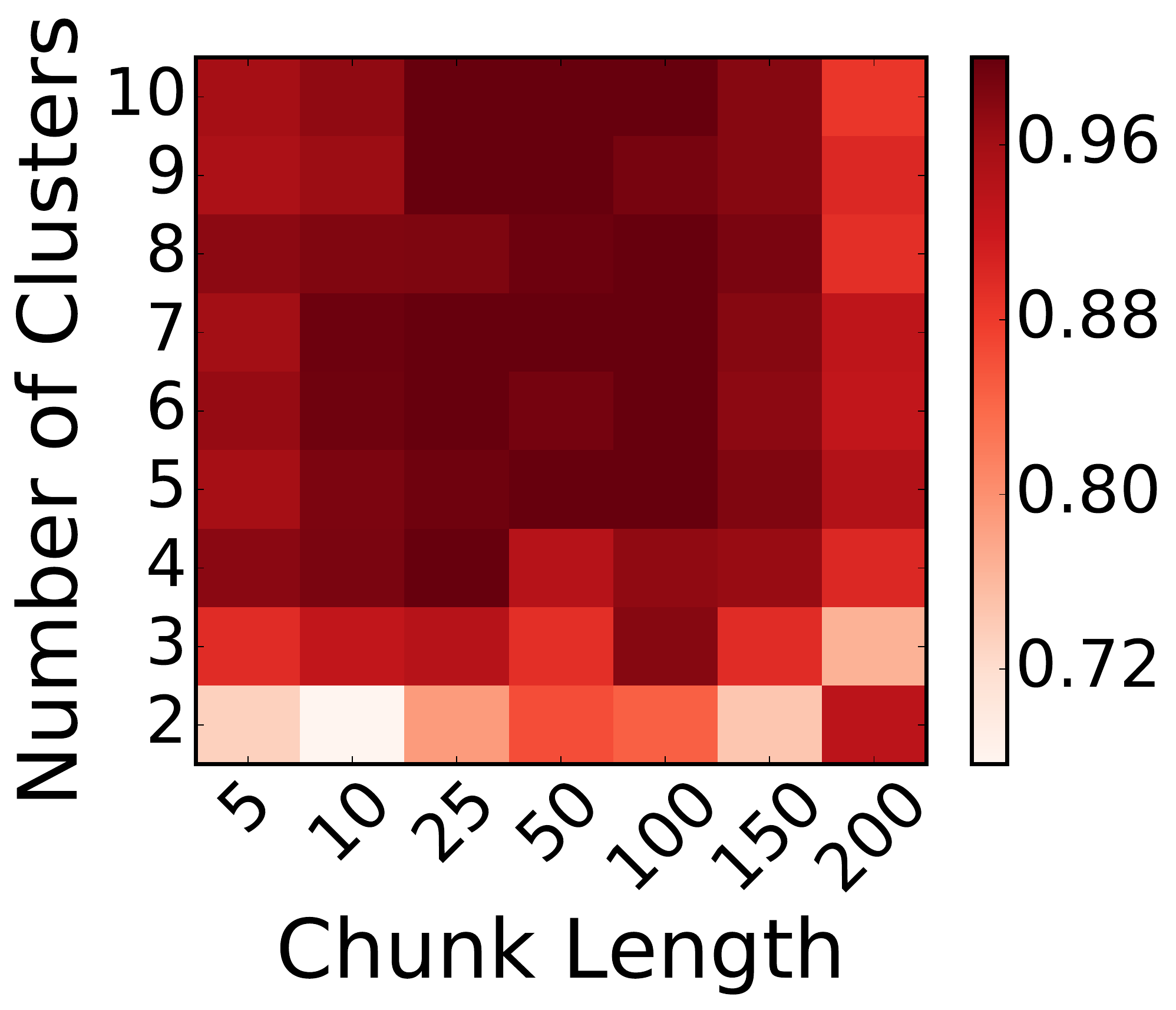}}&
			\hspace{-0.1in}	{\includegraphics[width=.15\textwidth]{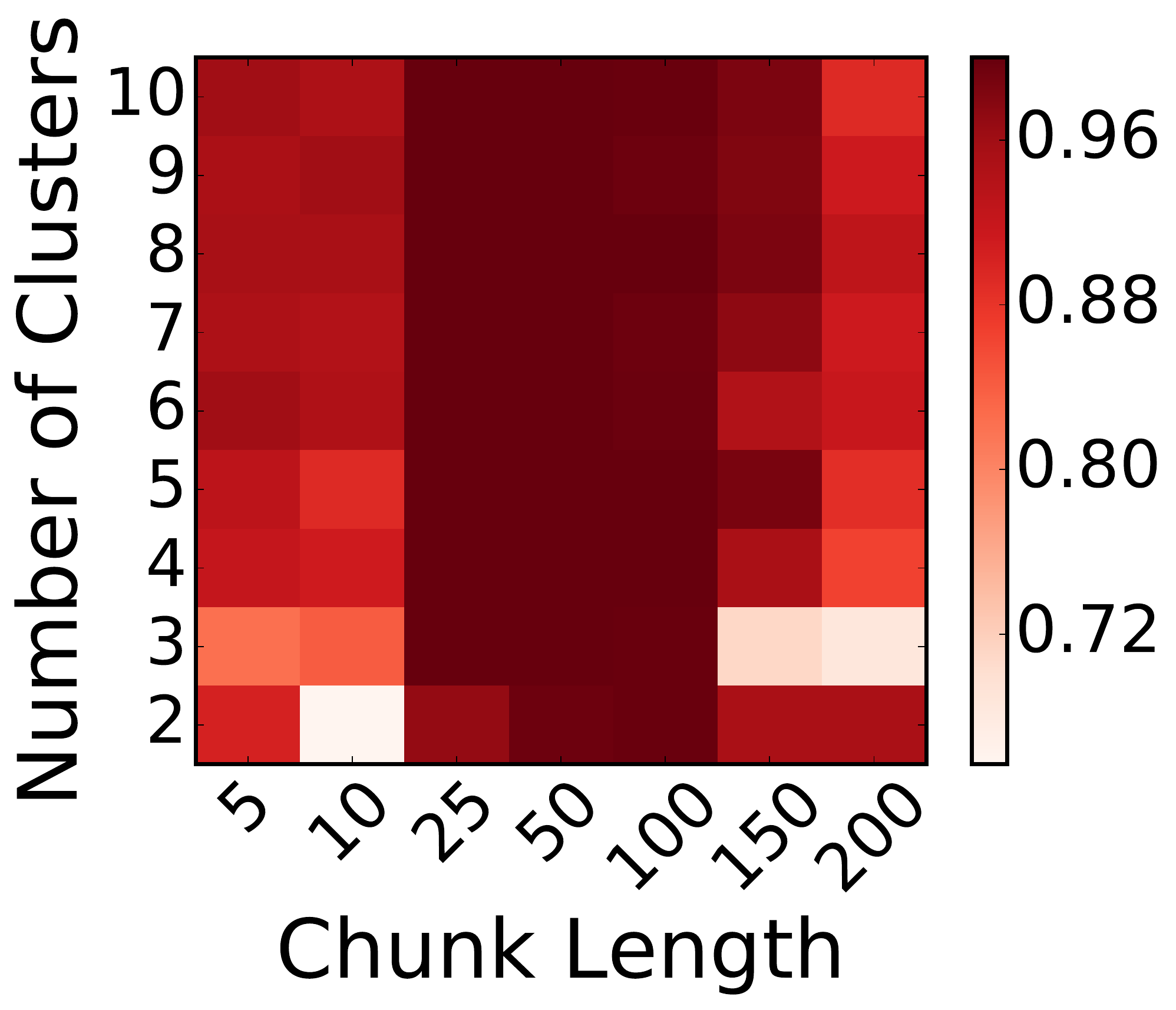}}\\
			(a) \DA & (b) \DB & (c) \DC 
		\end{tabular}
	\vspace{-4mm}
	\caption{Average precision for different chunk-length $C$ and number of clusters $K$. ($p = 75\%$)}
	\label{fig:clusobj}
	\vspace{-1mm}
\end{figure}

To validate our intuition for picking $C$, Figure \ref{fig:clusobj} shows a heatmap of the average precision obtained after clustering and anomaly-ranking on the test data (the attack and remaining 25\% benign graphs), for varying $C$ and $K$. We can see that for our chosen $C$ and $K$, \method~achieves near-ideal performance. We also find that the average precision appears robust to the number of clusters when chunk length is chosen in the ``safe region''.

\begin{figure} [h!]
	\vspace{-0.05in}
	\centering
	\begin{tabular}{ccc}
		\hspace{-0.1in}{\includegraphics[width=.33\columnwidth]{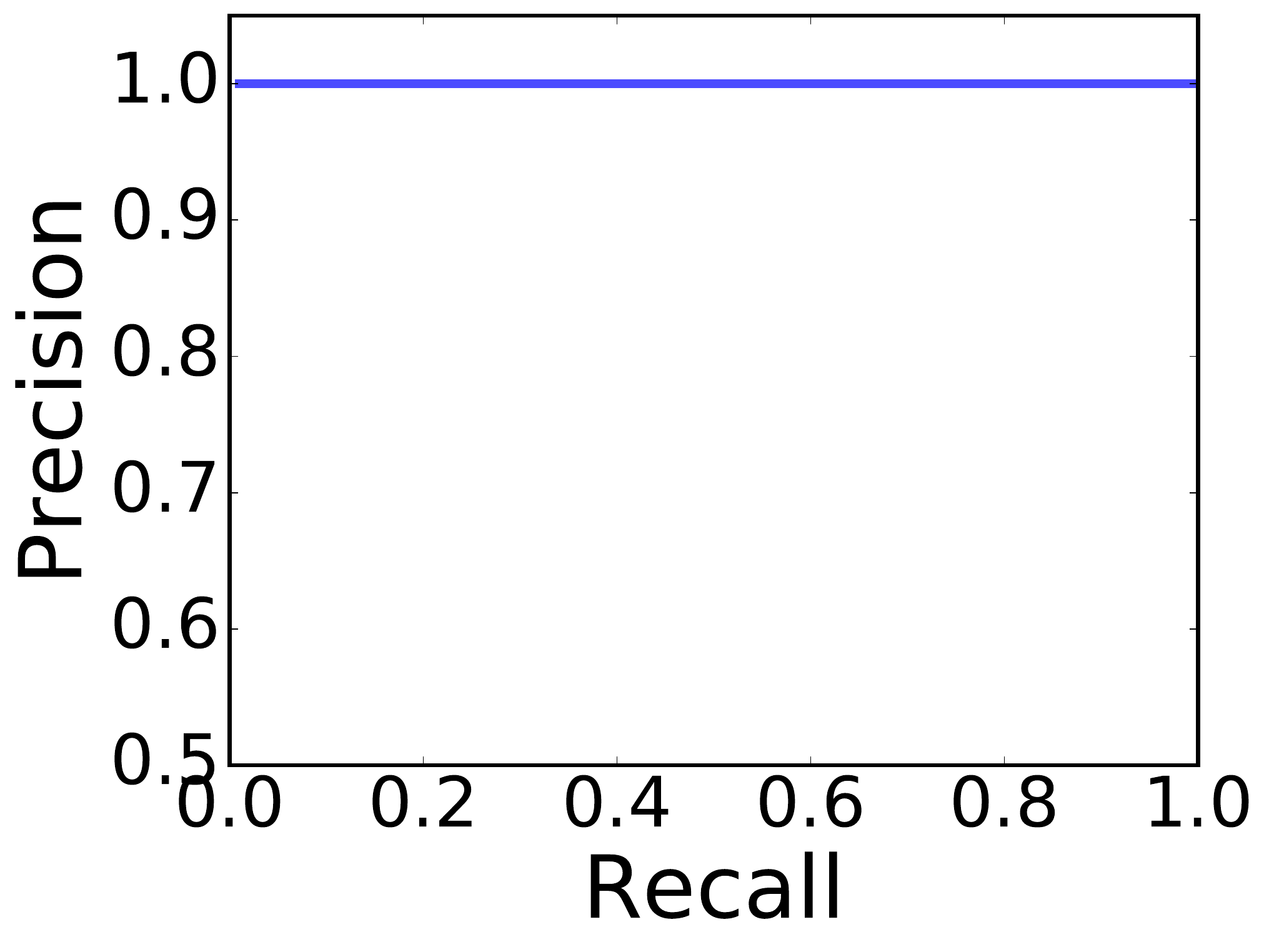}}&
		\hspace{-0.15in}{\includegraphics[width=.33\columnwidth]{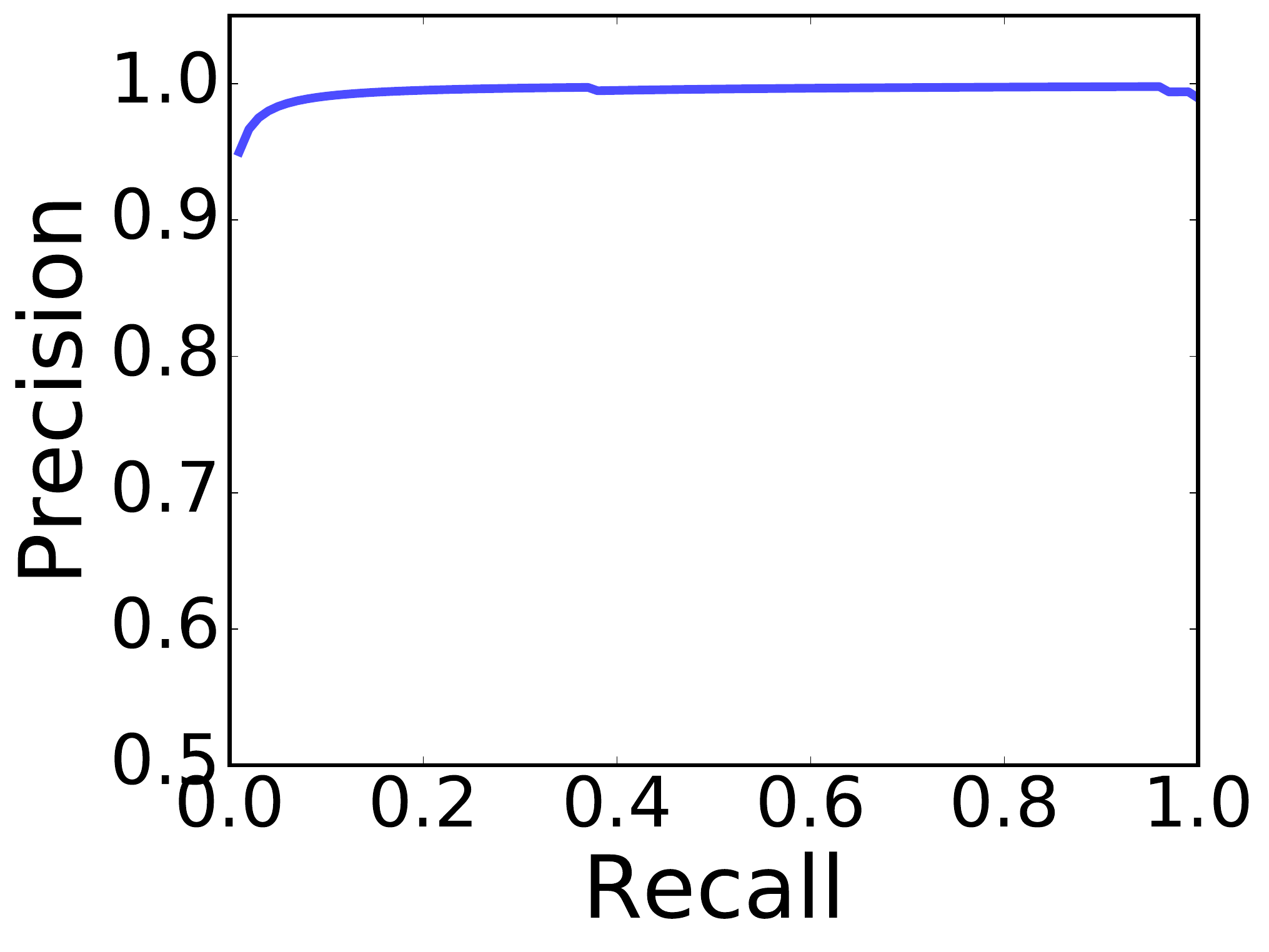}}&
		\hspace{-0.15in}{\includegraphics[width=.33\columnwidth]{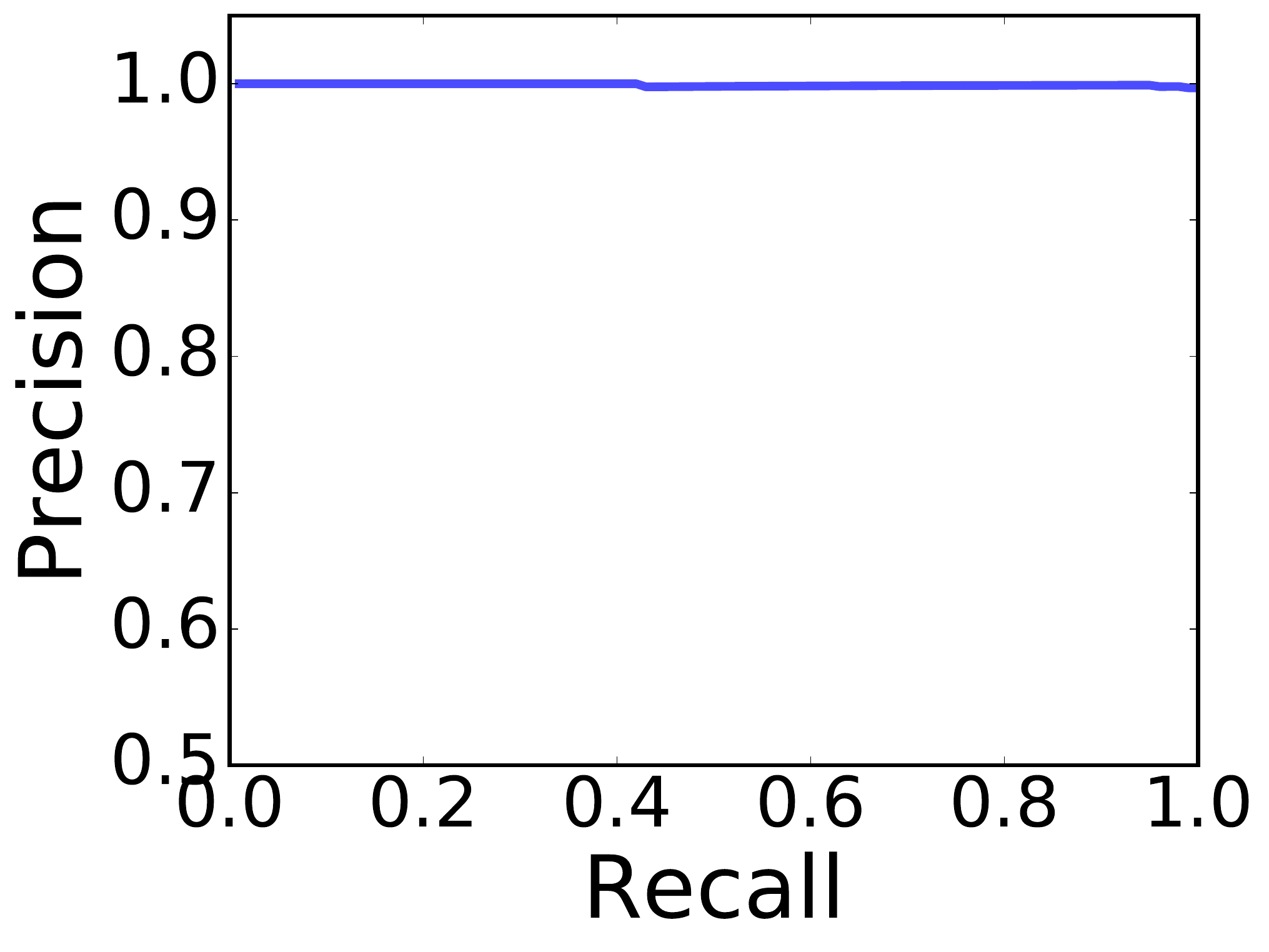}}\\
		\hspace{-0.1in}	{\includegraphics[width=.33\columnwidth]{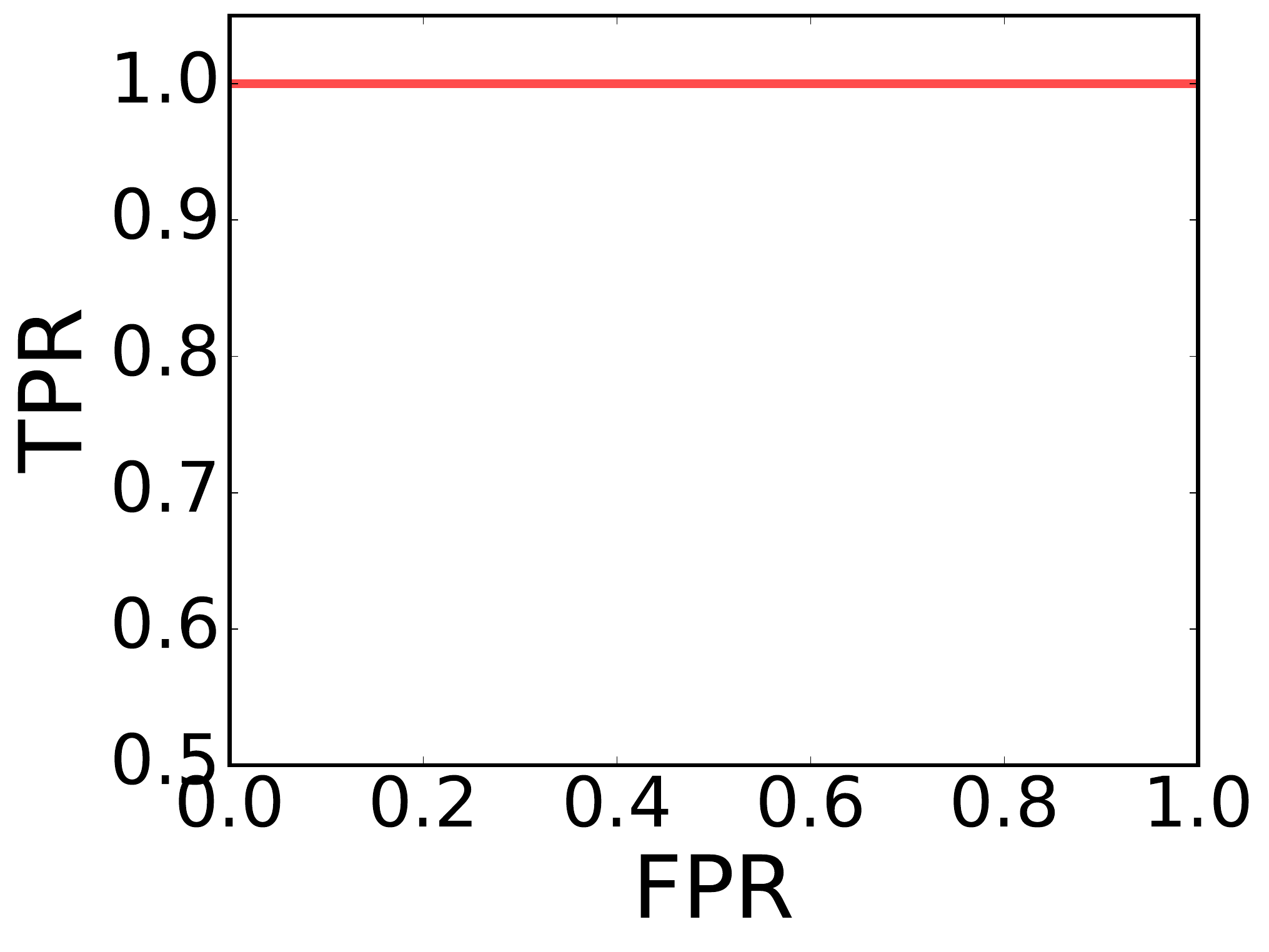}}&
		\hspace{-0.15in}	{\includegraphics[width=.33\columnwidth]{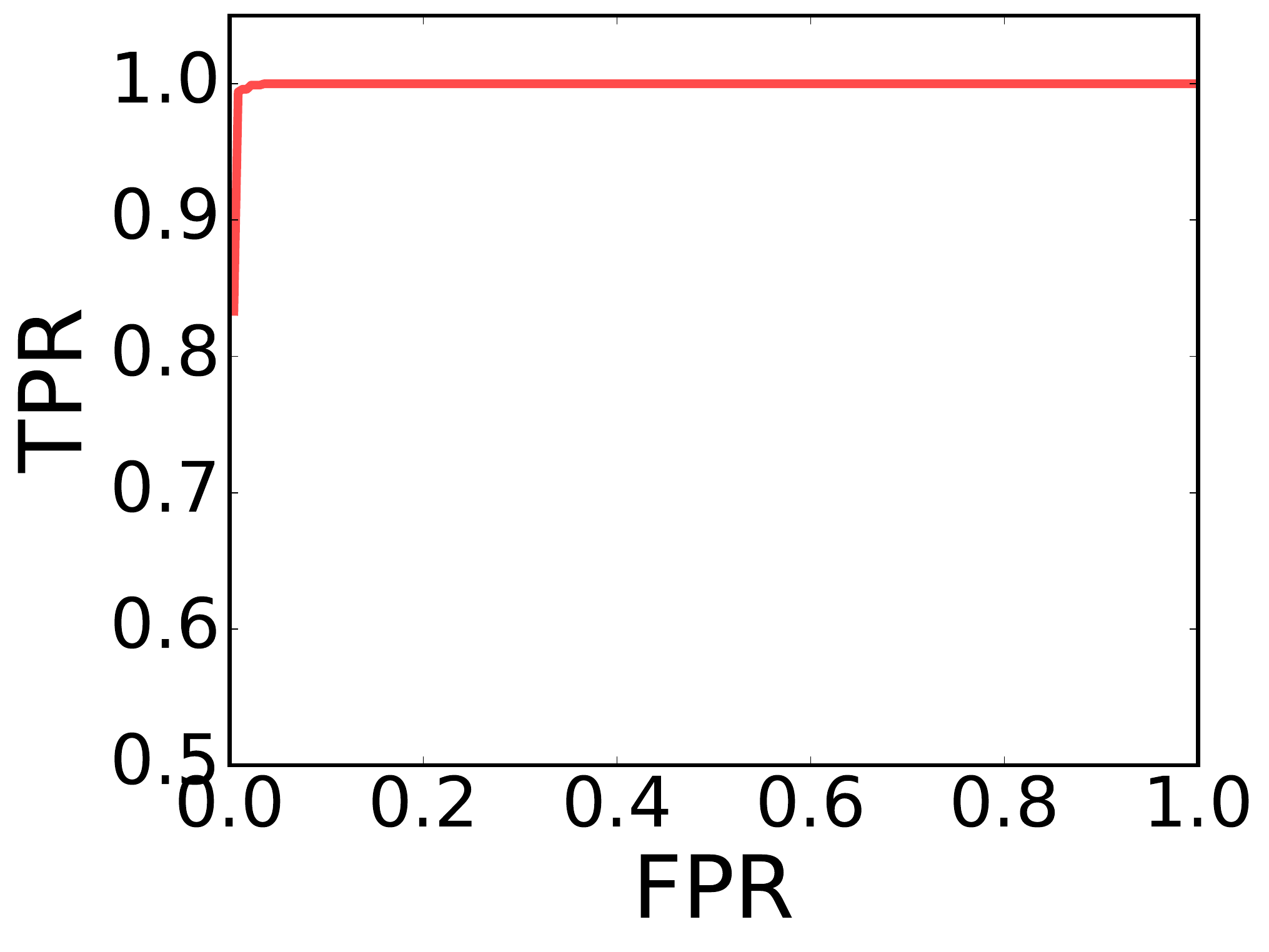}}&
		\hspace{-0.15in}	{\includegraphics[width=.33\columnwidth]{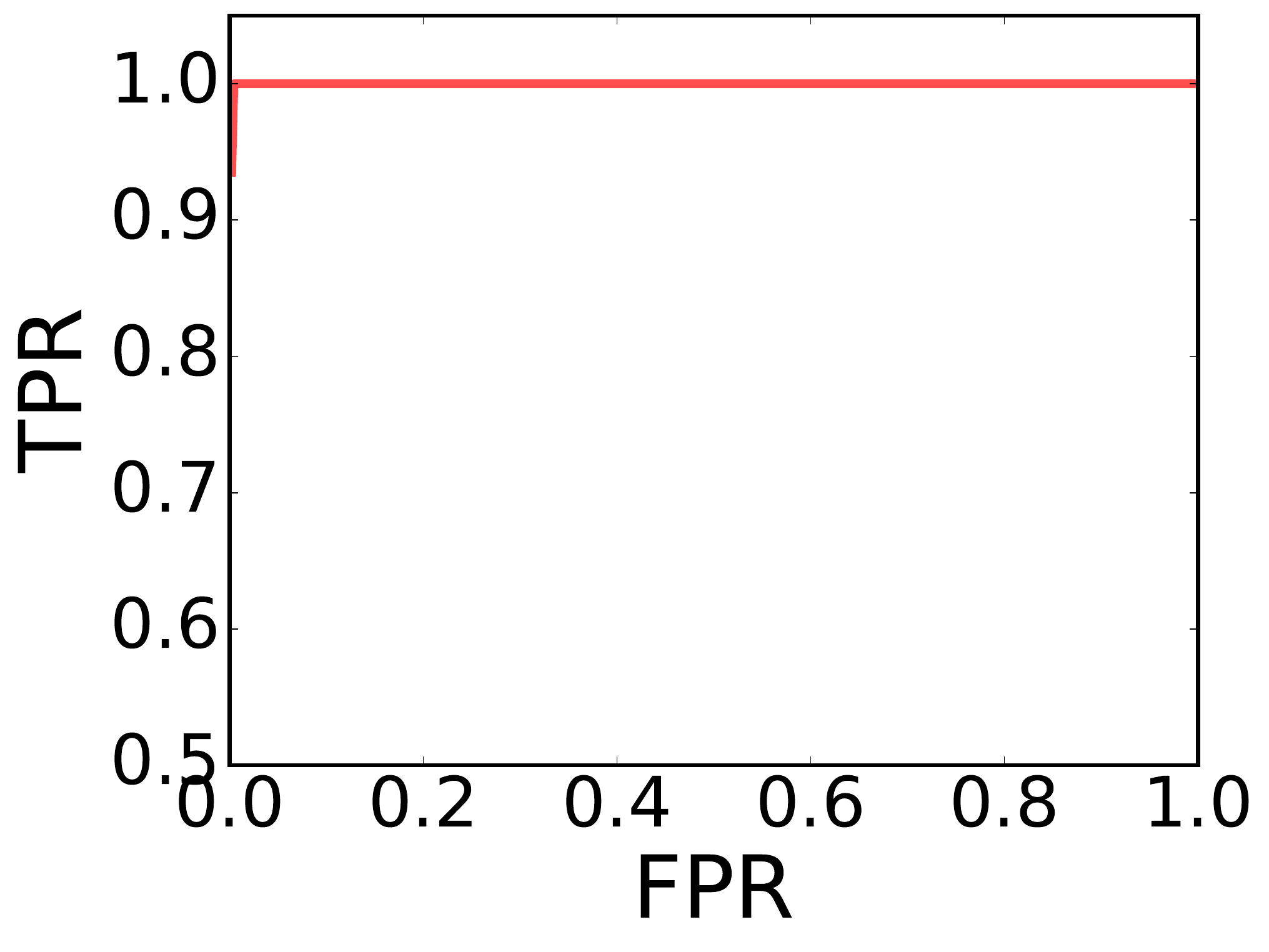}}\\
		(a) \DA & (b) \DB & (c) \DC \\
	\end{tabular}
	\vspace{-4mm}
	\caption{(top) Precision-Recall (PR) and (bottom) ROC curves averaged over 10 samples. ($p=25\%$)}
	\label{fig:ap}
	\vspace{-1mm}
\end{figure}

To quantify anomaly detection performance in the static setting, we set $p=25$\% of the data as training and cluster the training graphs based on their shingle vectors, following the aforementioned steps to choose $C$ and $K$. We then score each test graph by its distance to the closest centroid in the clustering. This gives us a ranking of the test graphs, based on which we plot the precision-recall (PR) and ROC curves. The curves (averaged over 10 independent random samples) for all the datasets are shown in Figure \ref{fig:ap}. We observe that even with 25\% of the data, static \method~ is effective in correctly ranking the attack graphs and achieves an average precision (AP,~area under the PR curve) of more then 0.9 and a near-ideal AUC (area under ROC curve).

%

\begin{figure}[h!]
	\vspace{-0.025in}
	\centering
			\begin{tabular}{ccc}
				\hspace{-0.1in}	{\includegraphics[width=.15\textwidth]{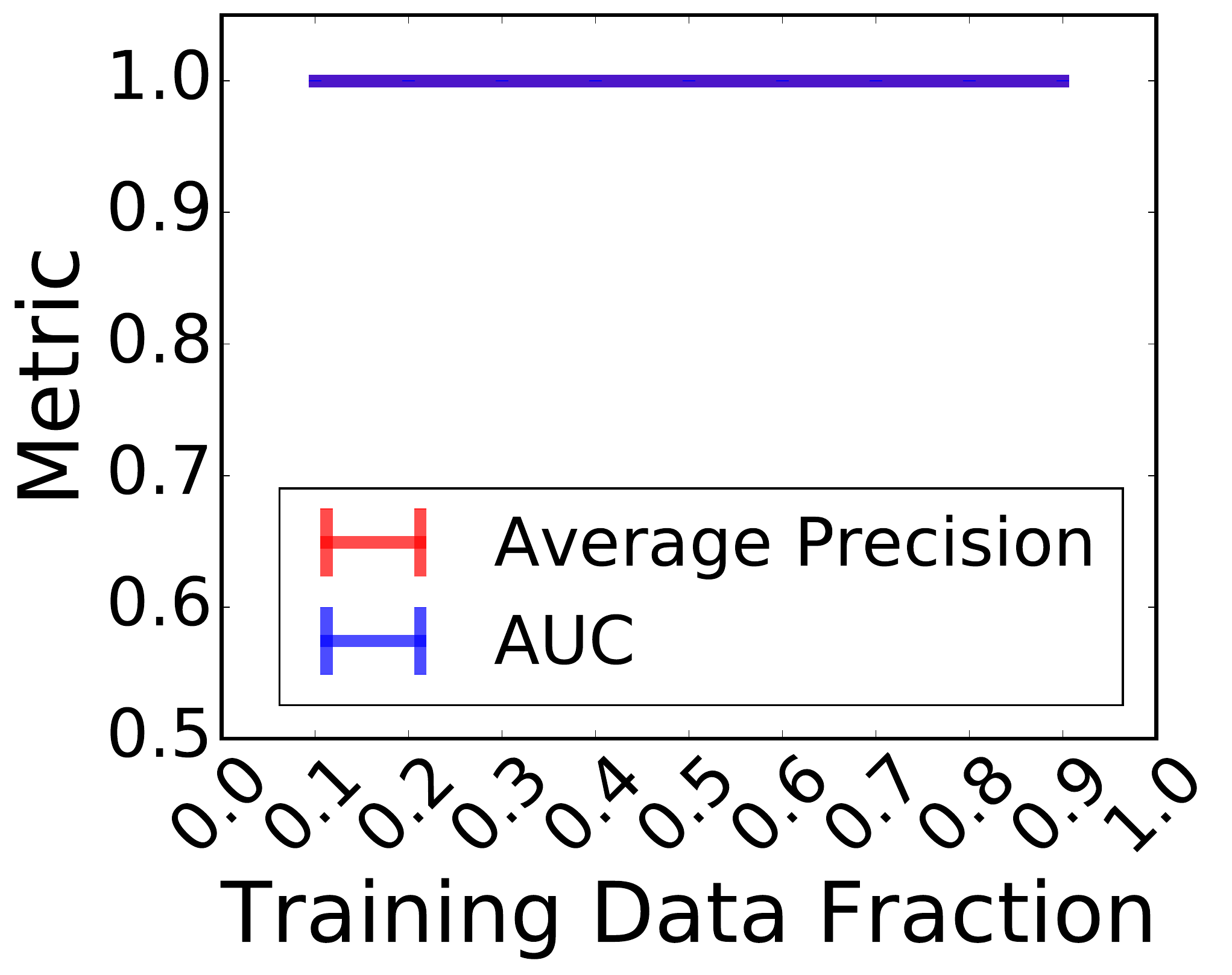}}&
				\hspace{-0.05in}	{\includegraphics[width=.15\textwidth]{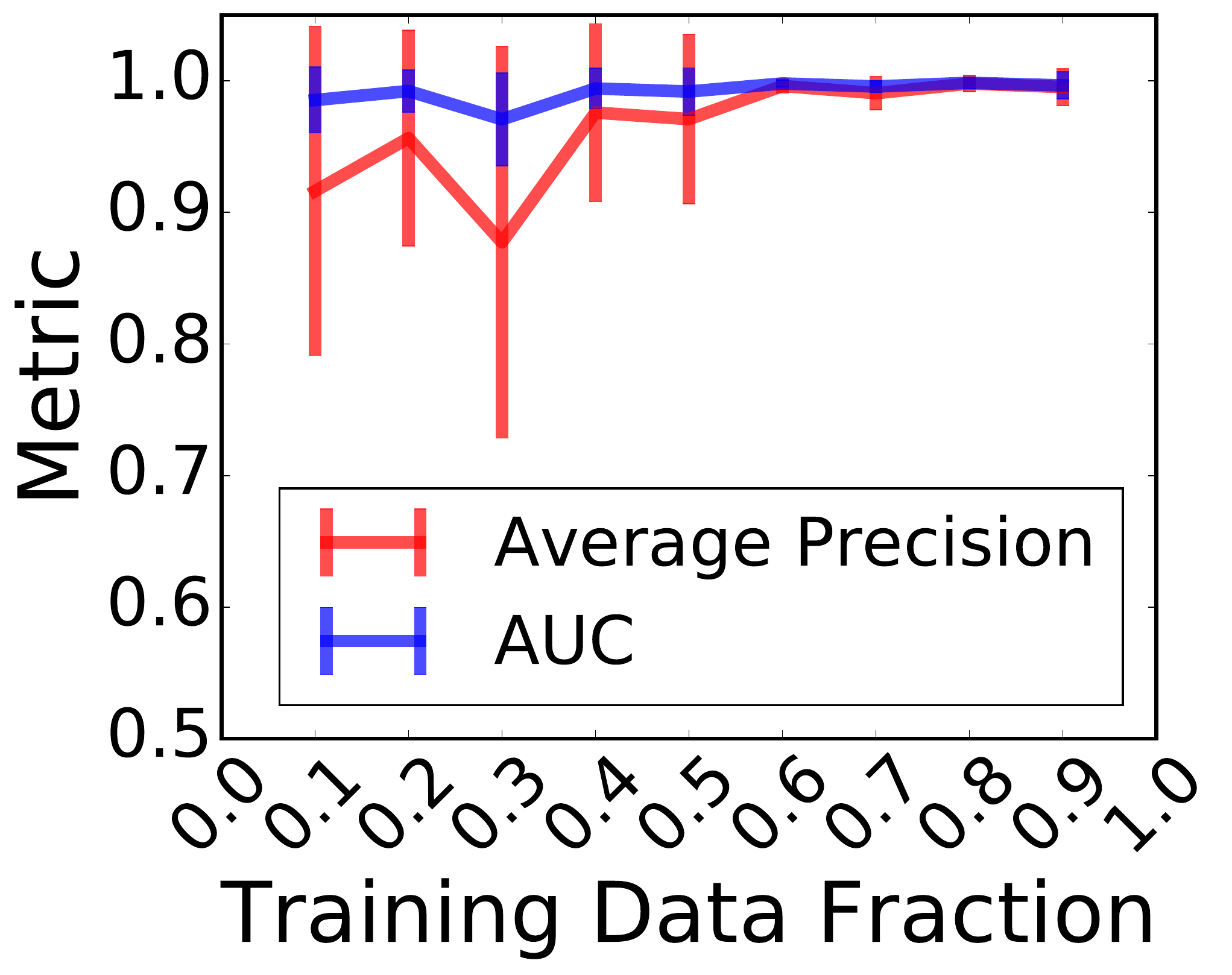}}&
				\hspace{-0.1in}	{\includegraphics[width=.15\textwidth]{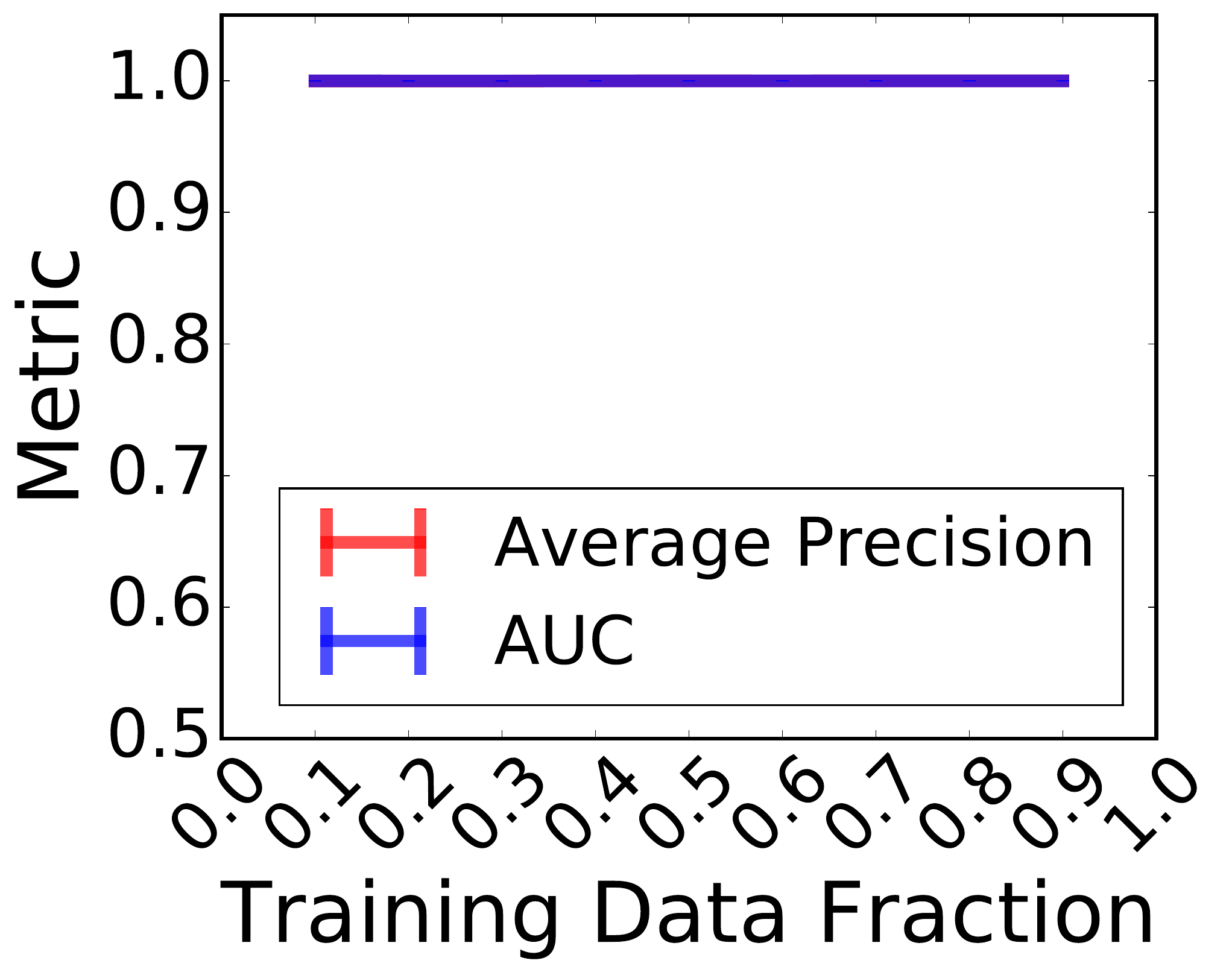}}\\
				(a) \DA & (b) \DB & (c) \DC 
			\end{tabular}
	\vspace{-4mm}
	\caption{AP and AUC with varying training \% $p$.}
	\label{fig:per}
	\vspace{-1mm}
\end{figure}

Finally, in Figure \ref{fig:per} we show how the AP and AUC change as the training data percentage $p$ is varied from $p=10$\% to $p=90$\%. We note that with sufficient training data, the test performance reaches an acceptable level for \DB.

The results in this section demonstrate a proof of concept that our proposed method can effectively spot anomalies provided offline data and unbounded memory. We now move on to testing \method~ in the streaming setting, for which it was designed.

\subsection{Streaming Evaluation}

\begin{figure*}[!t]
	\centering
			\begin{tabular}{ccc}
				{\includegraphics[width=.3\textwidth]{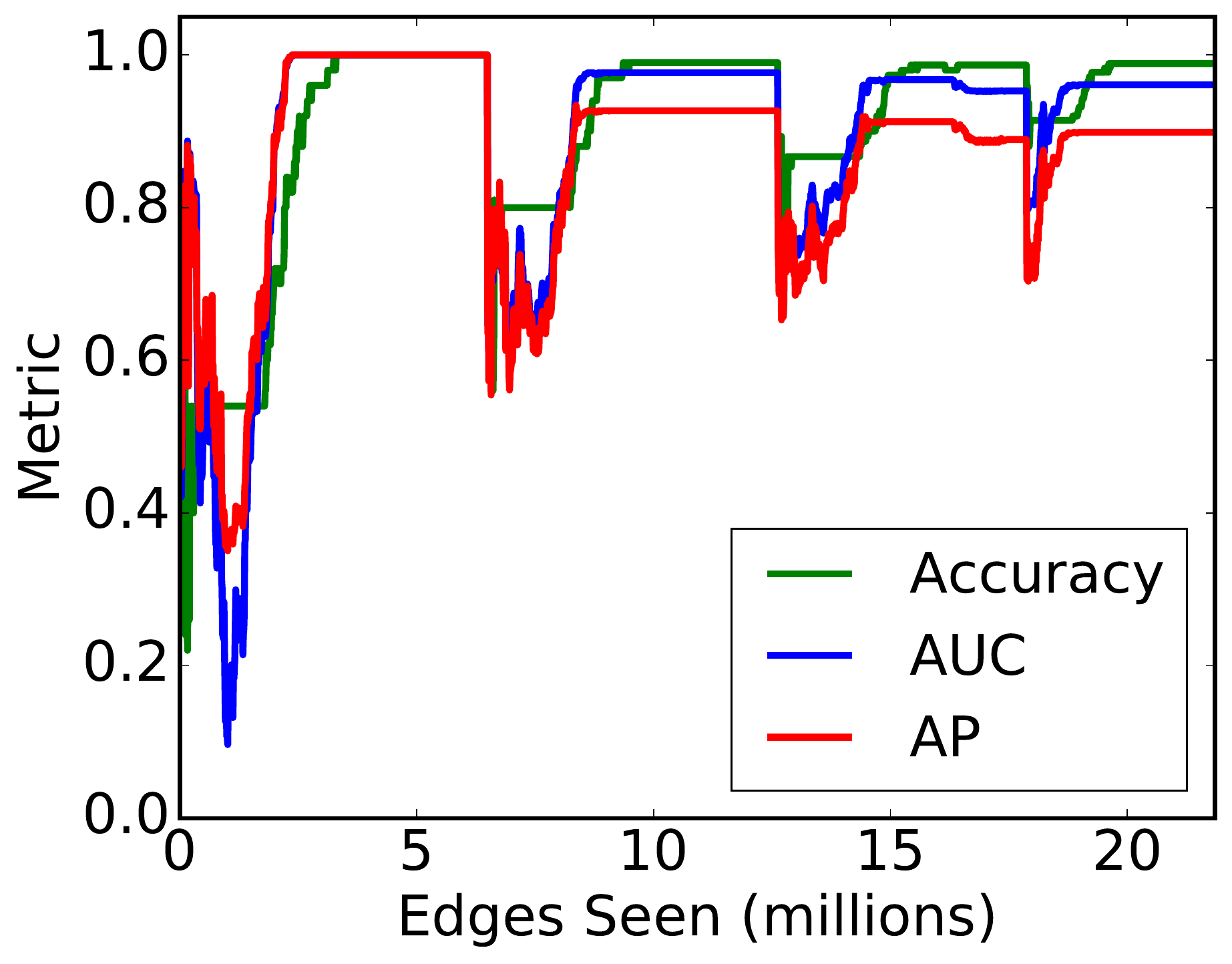}}&
				{\includegraphics[width=.3\textwidth]{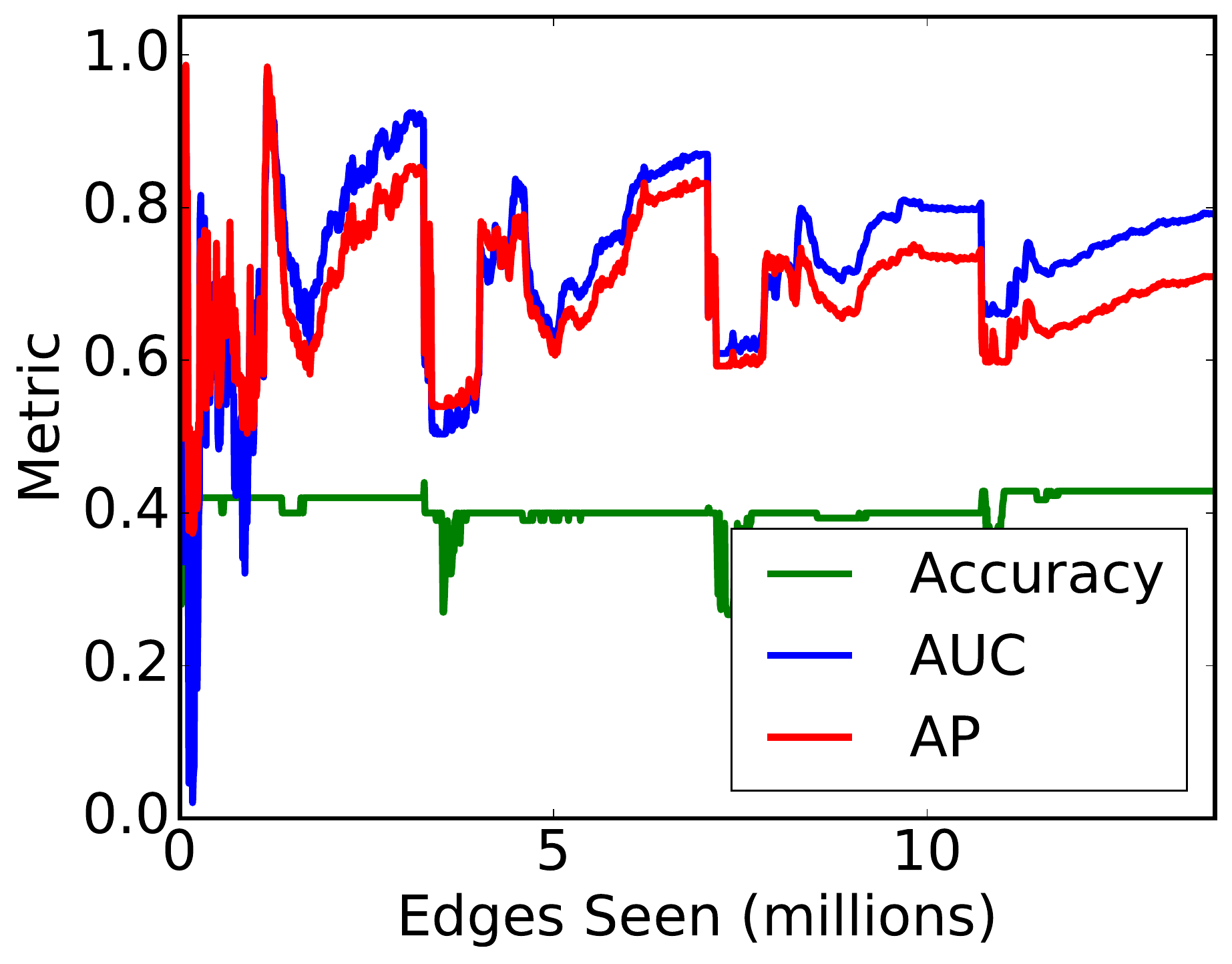}}&
				{\includegraphics[width=.3\textwidth]{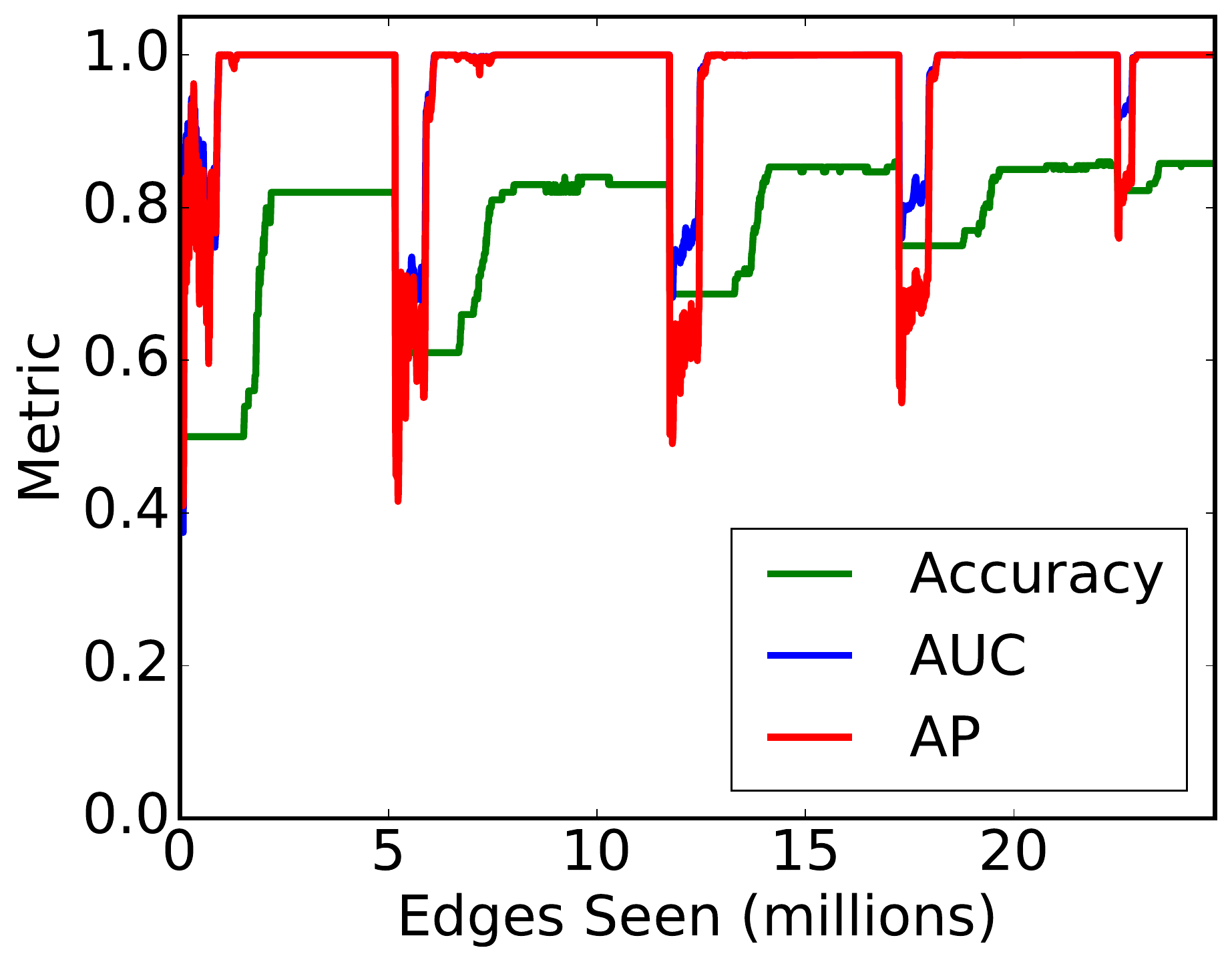}}\\
				(a) \DA & (b) \DB & (c) \DC 
			\end{tabular}
	\vspace{-2mm}
	\caption{Performance of \method~at different instants of the stream for all datasets ($L = 1000$).
	}
	\label{fig:alldatatime}
	\vspace{-0.125in}
\end{figure*}

\begin{figure}[h]
	\centering
	\subfigure{\includegraphics[width=.5\columnwidth]{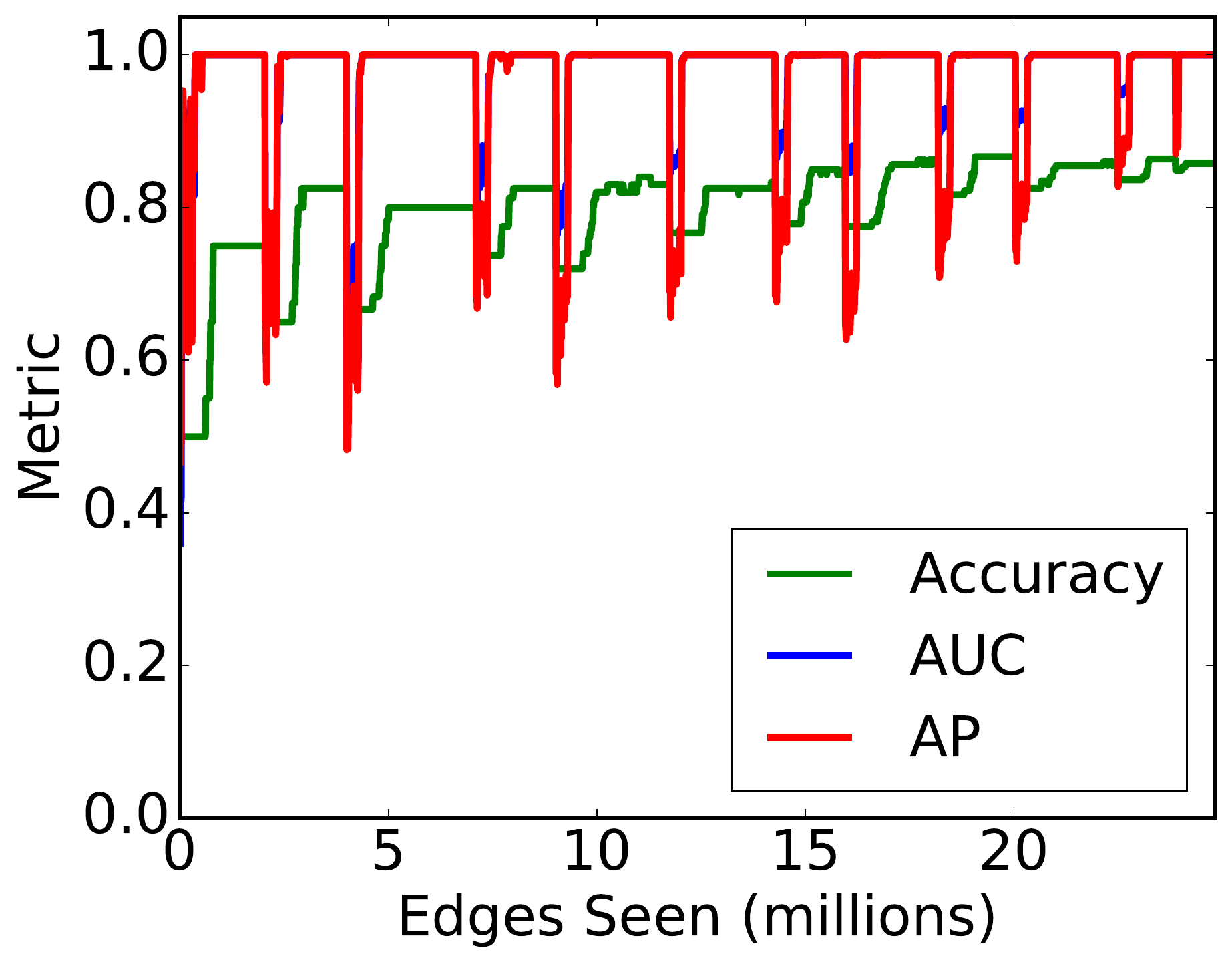}}\hfill
	\subfigure{\includegraphics[width=.5\columnwidth]{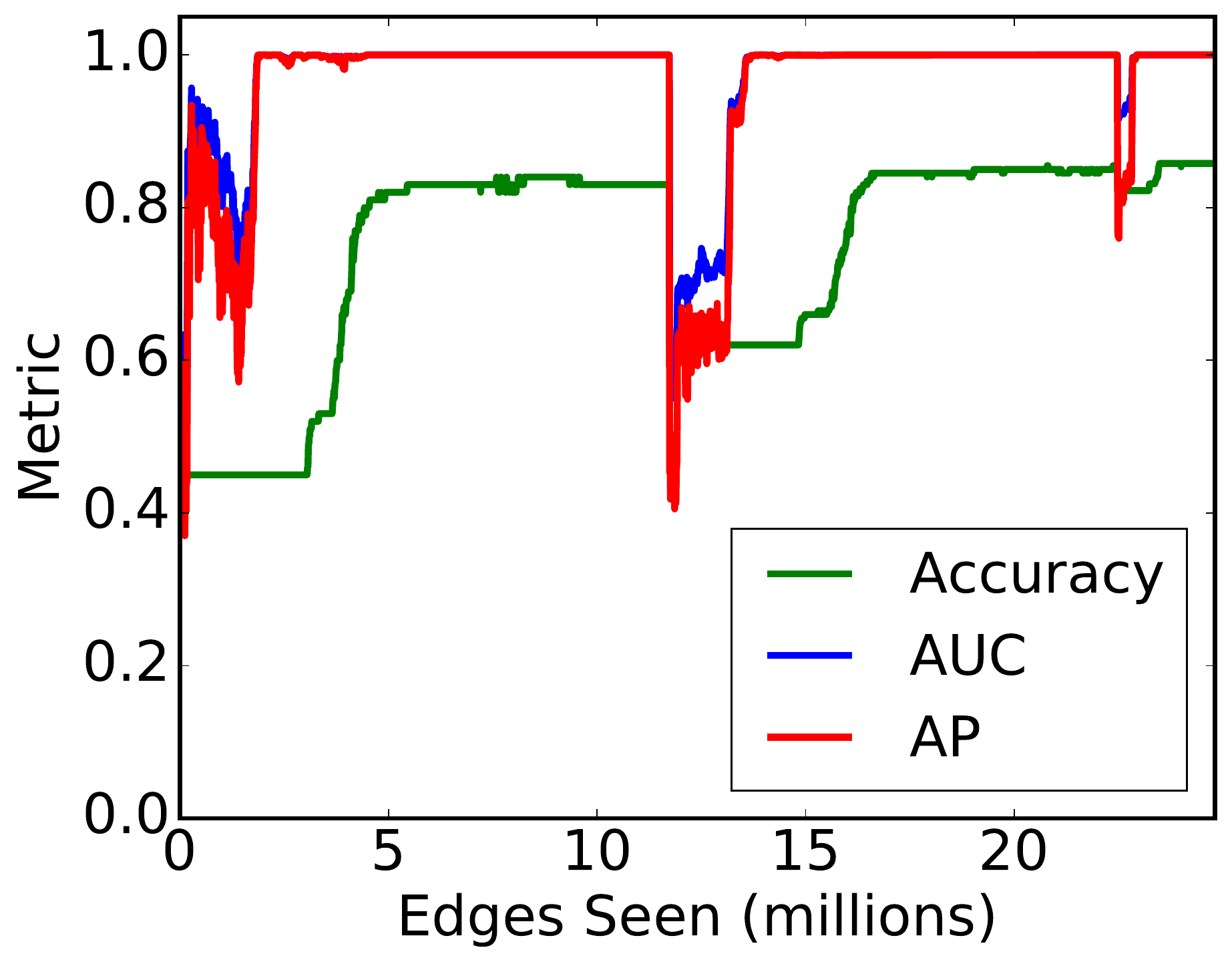}}
	\vspace{-6mm}
	\caption{\method~performance on \DC~ (measured at every 10K edges) when (left) $B=20$ graphs and (right) $B=100$ graphs arrive simultaneously.
		\label{fig:batchacc}}
	\vspace{-0.1in}
\end{figure}

We now show that \method~remains both accurate in detecting anomalies and efficient in processing time and memory usage in a streaming setting. We control the number of graphs that arrive and grow {\em simultaneously} with parameter $B$, by creating groups of $B$ graphs at random from the test graphs, picking one group at a time and interleaving the edges from graphs within the group to form the stream. In all experiments, 75\% of the benign graphs where used for bootstrap clustering. Performance metrics are computed on the instantaneous anomaly ranking every 10,000 edges.

\textbf{Detection performance.~}
Figure \ref{fig:batchacc} shows the anomaly detection performance on the \DC~dataset, when $B=20$ graphs grow in memory simultaneously (left) as compared to $B=100$ graphs (right). We make the following observations:
($i$) The detection performance follows a trend, with periodic dips followed by recovery.
($ii$) Each dip corresponds to the arrival of a new group of graphs. Initially, only a small portion of the new graphs are available and the detection performance is less accurate. However, performance recovers quickly as the graphs grow; the steep surges in performance in Figure \ref{fig:batchacc} imply a small anomaly detection delay.
($iii$) The average precision after recovery indicates a near-ideal ranking of attack graphs at the top.
($iv$) The dips become less severe as the clustering becomes more `mature' with increasing data seen; this is evident in the general upward trend of the accuracy.
($v$) The accuracy loss is not due to the ranking (since both AP and AUC remain high) but due to the chosen anomaly thresholds derived from the bootstrap clustering, where the error is due to false negatives.

Similar results hold for $B=50$ as shown in Figure \ref{fig:alldatatime} (c) for \DC, and (a) and (b) for \DA~and \DB~ respectively.

%
%
%
%
%
%
%
%
%

\begin{figure}[!t]
	\centering
		\begin{tabular}{cc}
			\hspace{-0.2in}	{\includegraphics[width=.5\columnwidth]{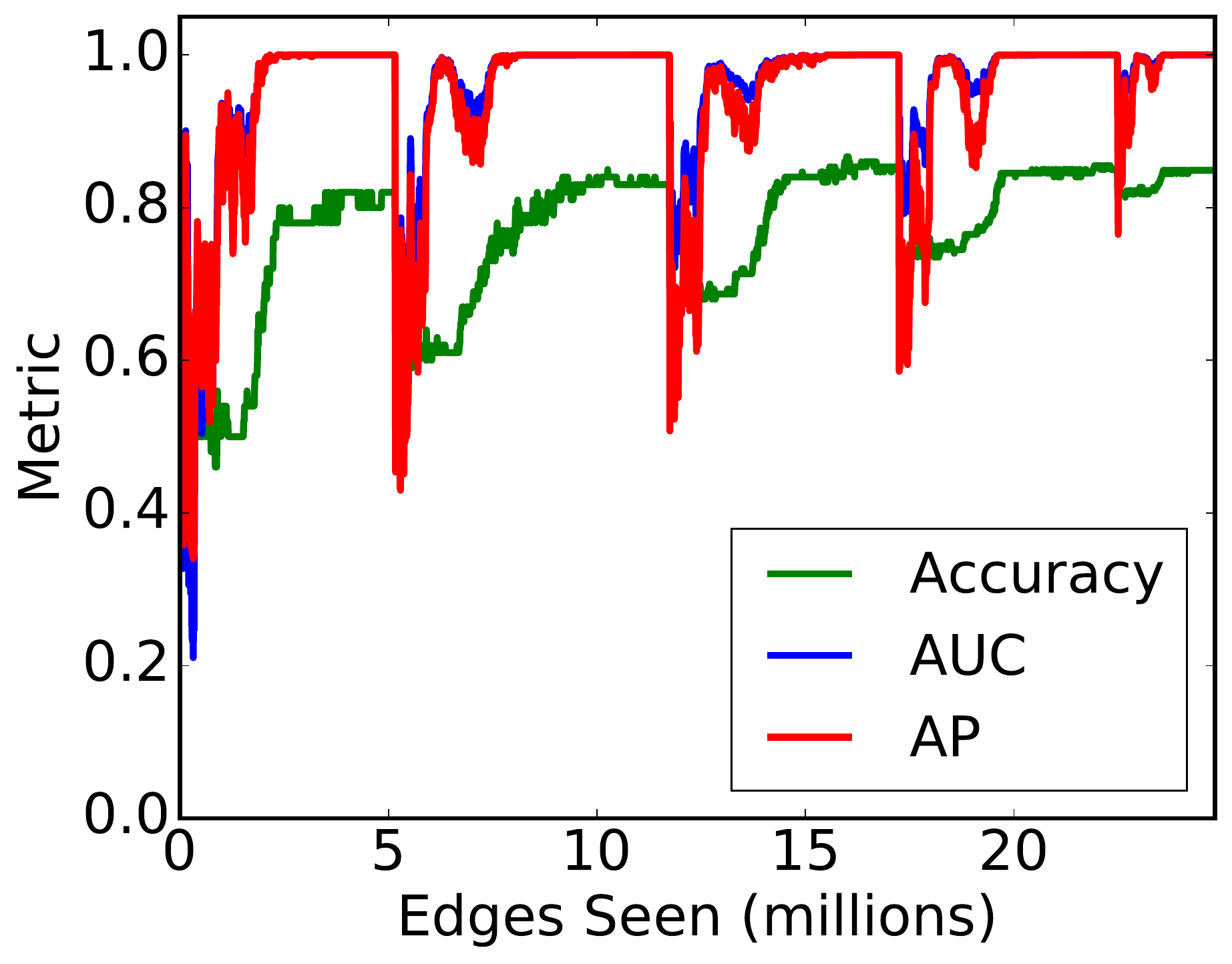}}&
		\hspace{-0.1in}		{\includegraphics[width=.5\columnwidth]{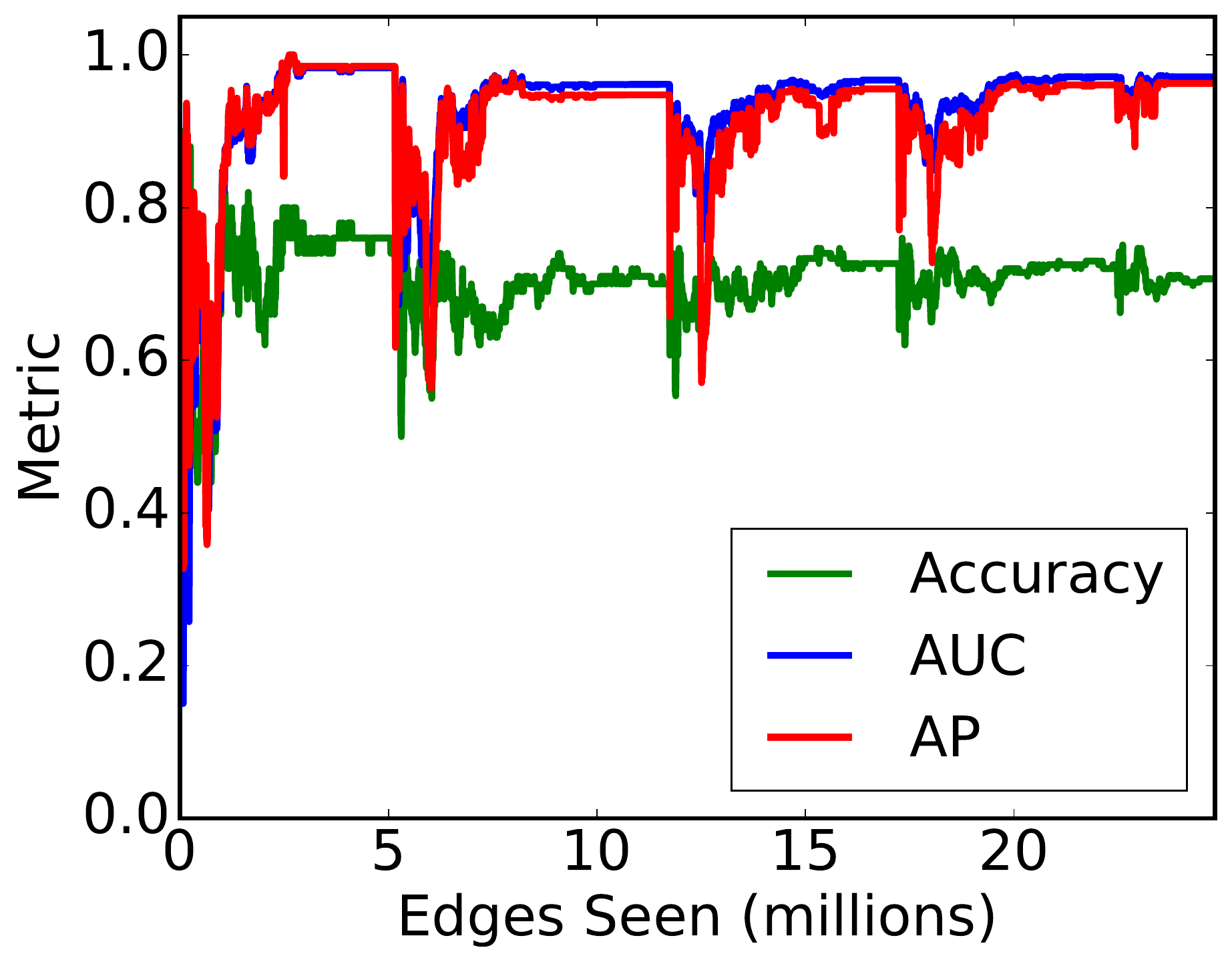}}\\
			(a) $L = 100$ & (b) $L = 10$
		\end{tabular}
	\vspace{-3mm}
	\caption{Performance of \method ~on \DC ~for different values of the sketch size.}
	\label{fig:sketchsize}
	\vspace{-2mm}
\end{figure}

\textbf{Sketch size.~}
Figures \ref{fig:alldatatime} and \ref{fig:batchacc} show \method's performance for sketch size $L=1000$ bits. When compared to the number of unique shingles $|S|$ in each dataset (649,968, 580,909 and 1,106,684 for \DA, \DB~ and \DC~), sketching saves considerable space. Reducing the sketch size saves further space but increases the error of cosine distance approximation. Figure \ref{fig:sketchsize} shows \method's performance on \DC ~for smaller sketch sizes. Note that it performs equally well for $L=100$ (compared to Fig. \ref{fig:alldatatime}(c)), and reasonably well even with sketch size as small as $L=10$.

\begin{figure*}[!t]
	\centering
				\begin{tabular}{ccc}
					{\includegraphics[width=.3\textwidth]{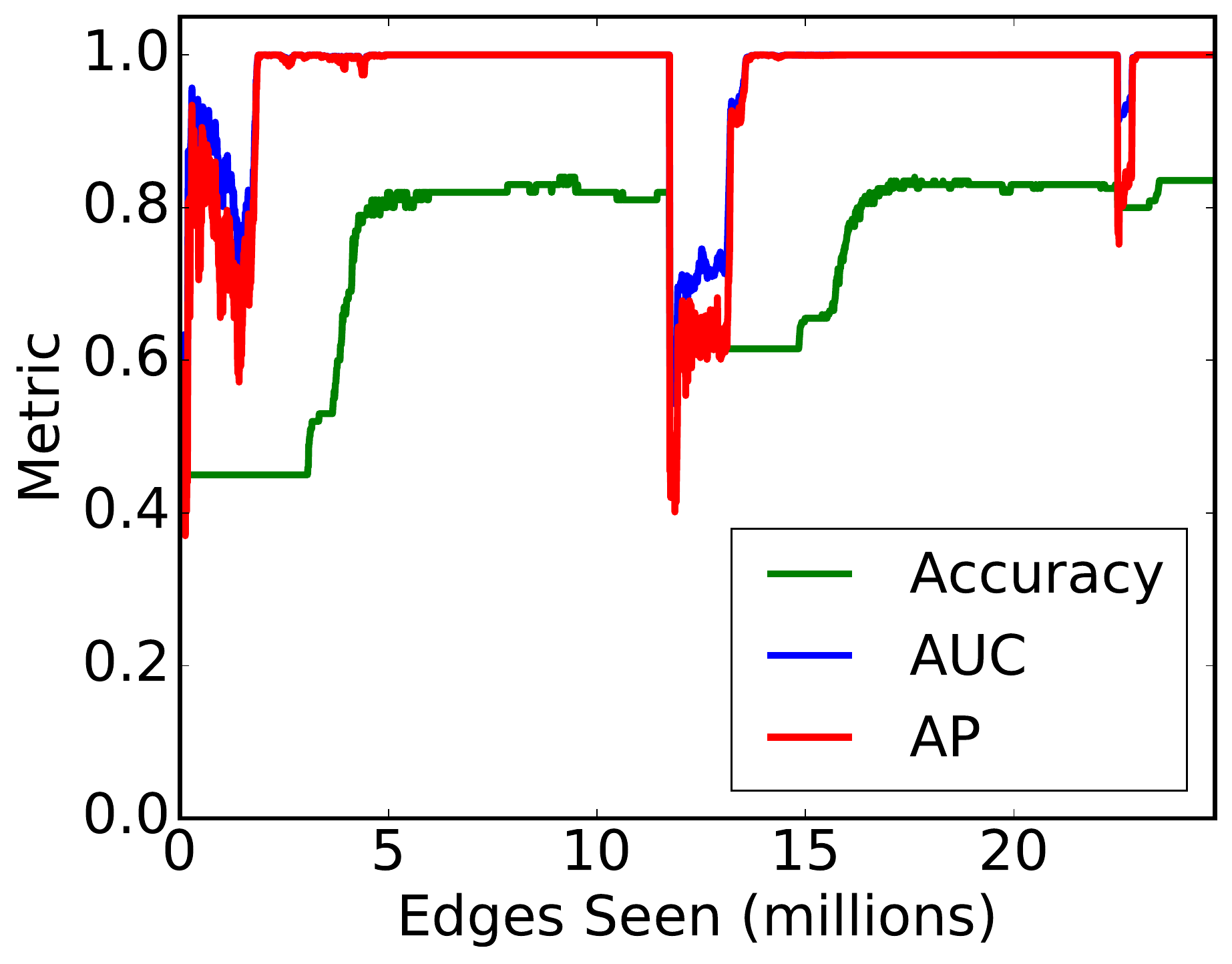}}&
					{\includegraphics[width=.3\textwidth]{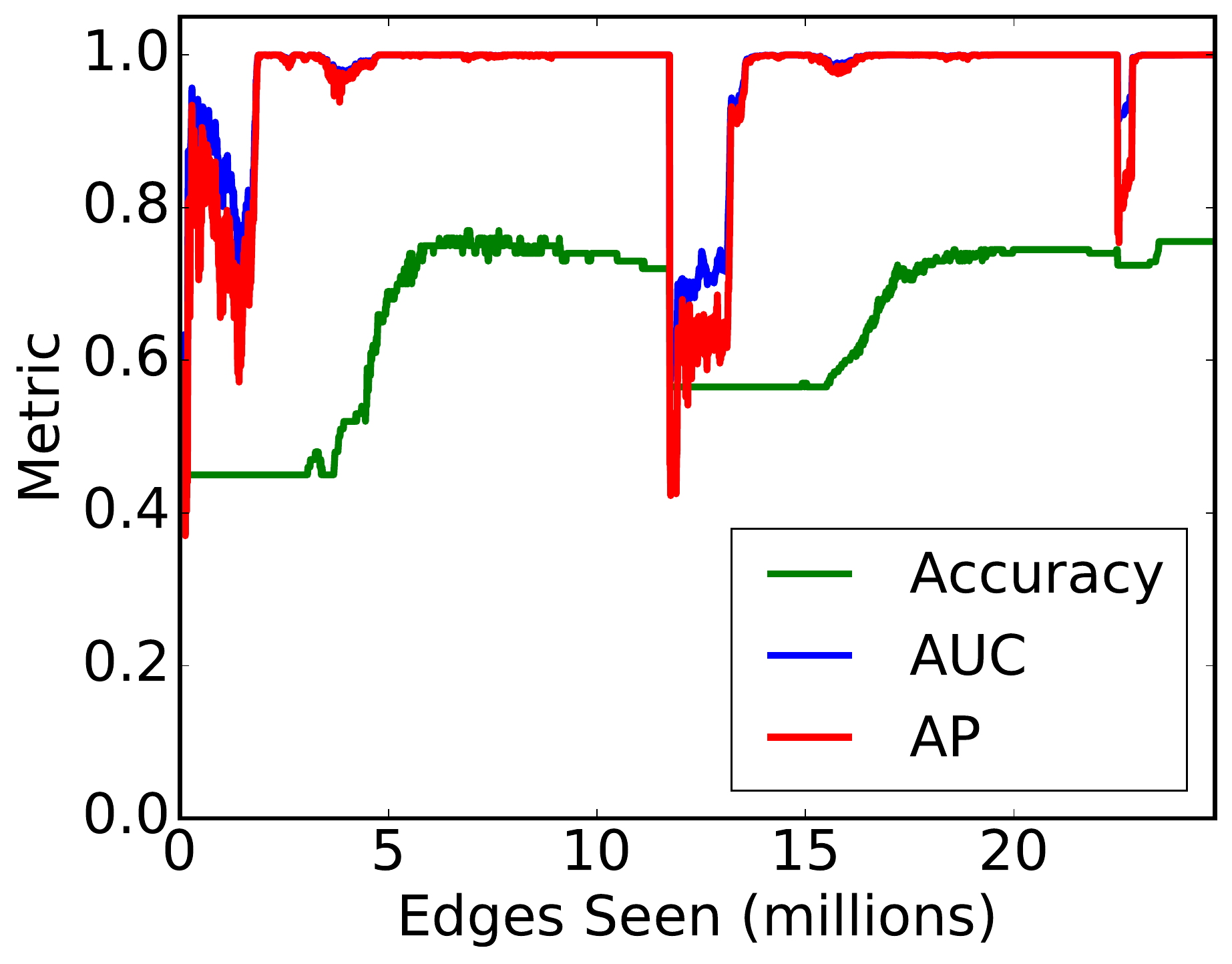}}&
					{\includegraphics[width=.3\textwidth]{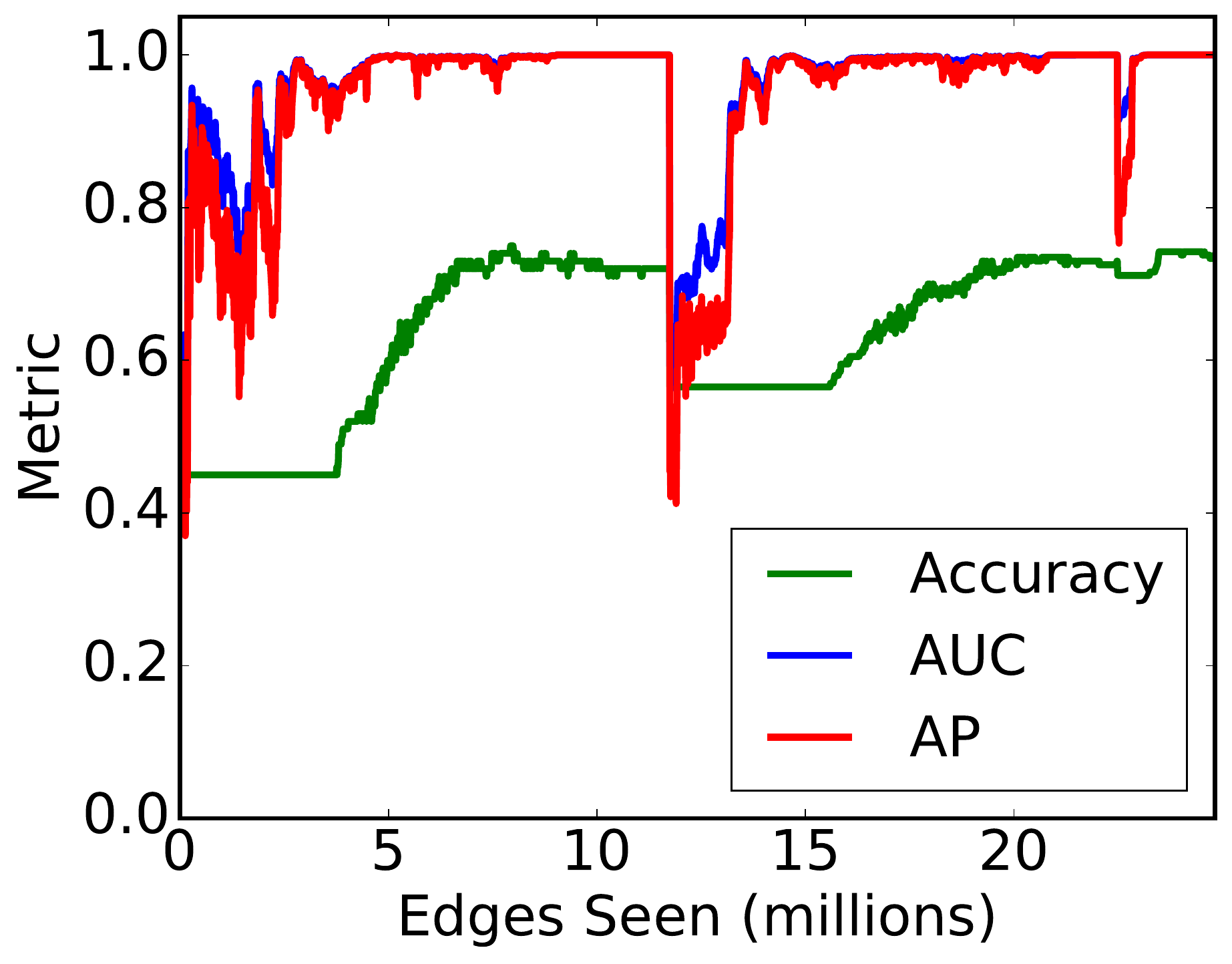}}\\
					(a) Limit = 15\% & (b) Limit = 10\% & (c) Limit = 5\%
				\end{tabular}
	\vspace{-2mm}
	\caption{Performance of \method ~on \DC ~($L = 1000$), for different values of the memory limit $N$ (as a fraction of the number of incoming edges).}
	\label{fig:memory}
	\vspace{-3mm}
\end{figure*}

\begin{figure}[!h]
	\centering
	\begin{tabular}{cc}
		\hspace{-0.125in}\includegraphics[width=.54\columnwidth]{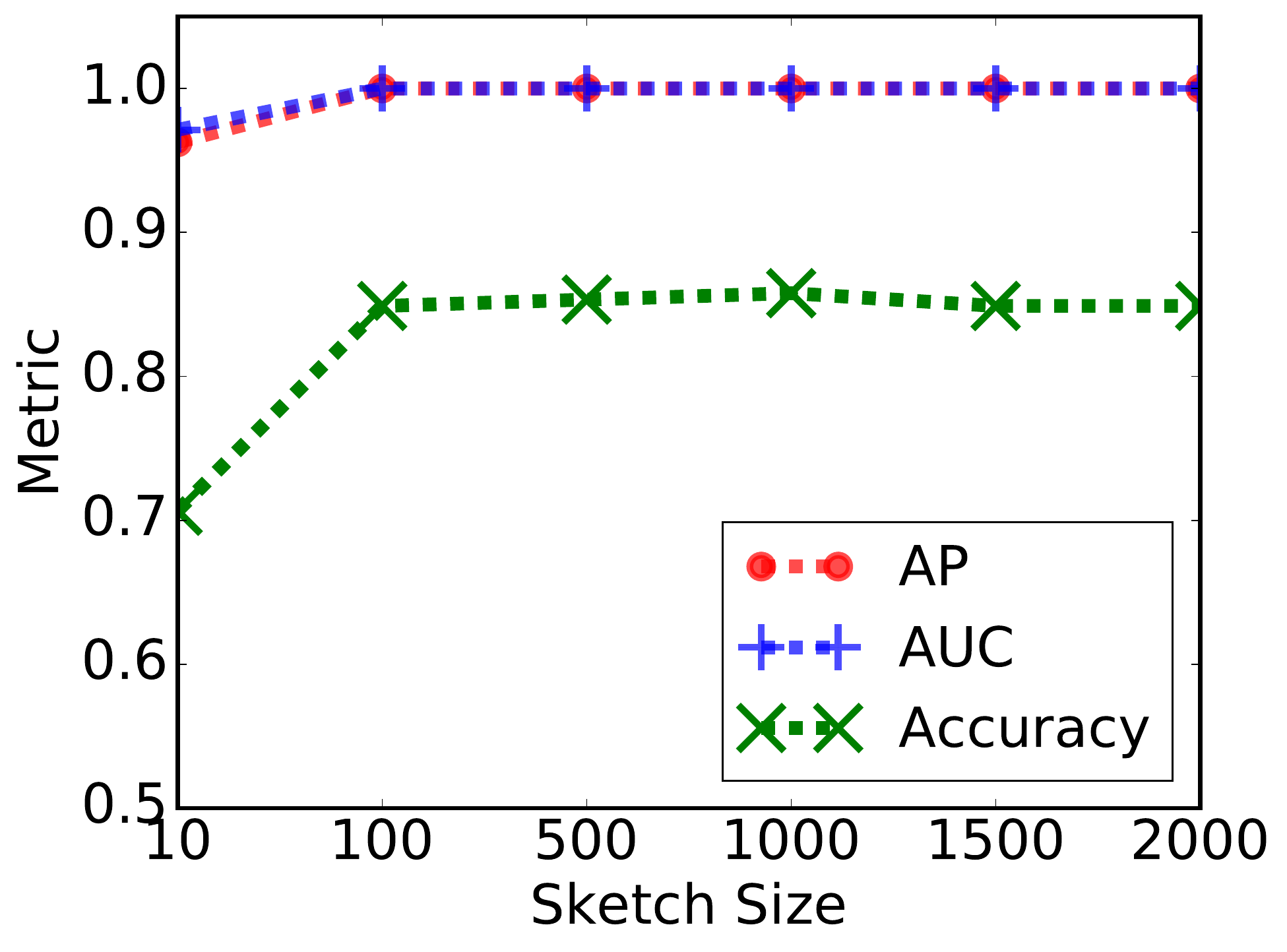} &
		\hspace{-0.2in}\includegraphics[width=.54\columnwidth]{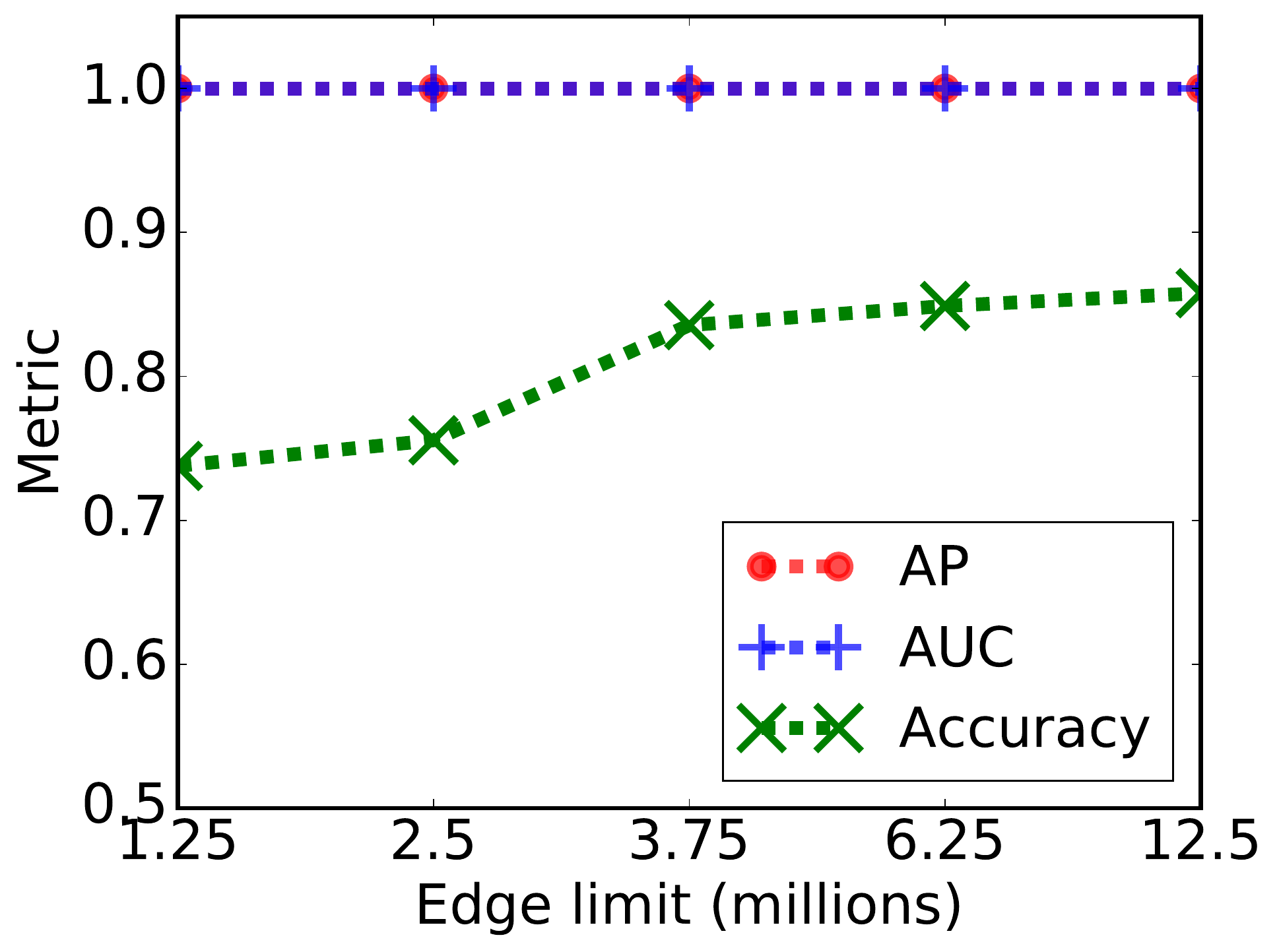}
	\end{tabular}
	\vspace{-5mm}
	\caption{\method~ performance on \DC~with increasing (left) sketch size $L$ and (right)
		memory limit $N$.}
	\label{fig:LN}
	\vspace{-2mm}
\end{figure}

\textbf{Memory limit.~}
Now we investigate \method's performance by limiting the memory usage using $N$: the maximum number of allowed edges in memory at any given time. Figures \ref{fig:memory} (a)--(c) show the performance when $N$ is limited to 15\%, 10\%, and 5\% of the incoming stream of $\sim$25M edges, on \DC~(with $B=100$, $L=1000$). The overall performance decreases only slightly as memory is constrained. The detection delay (or the recovery time) increases, while the speed and extent of recovery decays slowly. This is expected, as with small memory and a large number of graphs growing simultaneously, it takes longer to observe a certain fraction of a graph at which effective detection can be performed.

These results indicate that \method~ continues to perform well even with limited memory. Figure \ref{fig:LN} shows performance at the end of the stream with (a) increasing sketch size $L$ ($N$ fixed at 12.5M edges) and (b) increasing memory limit $N$ ($L$ fixed at 1000). \method~ is robust to both parameters and demonstrates stable performance across a wide range of their settings.

\begin{figure}[h]
	\vspace{-0.1in}
	\centering
	\begin{tabular}{cc}
		\hspace{-0.15in}	\includegraphics[width=.29\textwidth,height=1.65in]{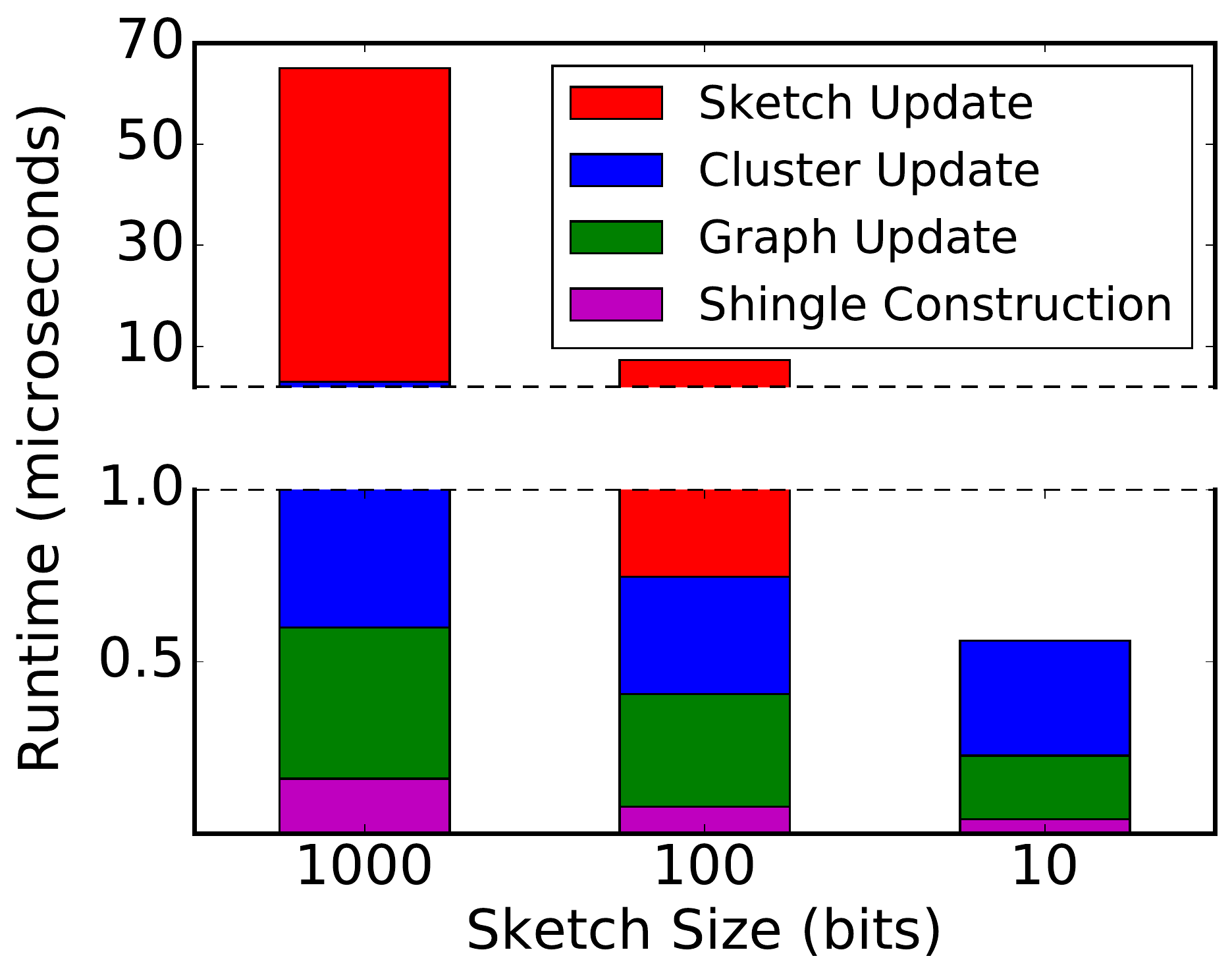} &
		\hspace{-0.15in}	\includegraphics[width=0.21\textwidth]{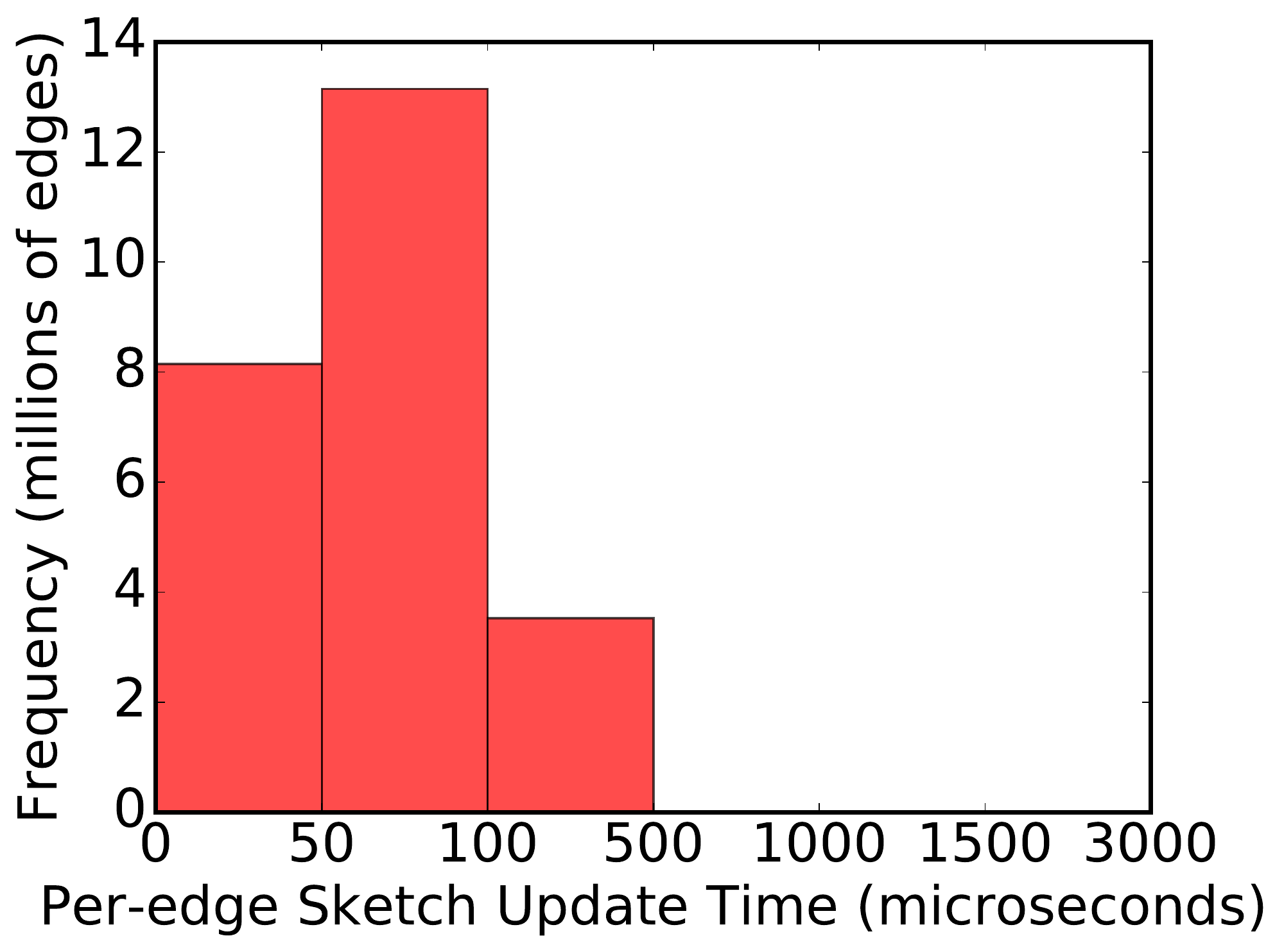}
	\end{tabular}
	\vspace{-5mm}
	\caption{(left) Average runtime of \method~per edge, for various sketch sizes on \DC.
		(right) Distribution of sketch processing times for $\sim$25M edges.
	}
	\label{fig:runtime}
	\vspace{-0.05in}
\end{figure}

\textbf{Running time.~}
To evaluate the scalability of \method~ for high-volume streams, we measure its per-edge running time on \DC~for sketch sizes $L=1000,100$ and $10$ averaged over the stream of $\sim$25M edges, in Figure \ref{fig:runtime} (left). Each incoming edge triggers four operations: updating the graph adjacency list, constructing shingles, updating the sketch/projection vector and updating the clusters. We observe that the running time is dominated by updating sketches (hashing strings), but the total running time per edge is under 70 microseconds when $L=1000$ (also see Fig. \ref{fig:runtime} (right)). Thus, \method~can scale to process more than 14,000 edges per second. For $L=100$, which produces comparable performance (see Fig. \ref{fig:sketchsize} (a)), it can further scale to over 100,000 edges per second.


\textbf{Memory usage.~}
Finally, we quantify the memory usage of \method. Table \ref{tbl:efficiency}
shows the total memory  consumed for \DC~by graph edges ($M_G$) in {\sc mb}'s and by projection and sketch vectors ($M_Y$, $M_X$) in {\sc kb}'s, with increasing $N$ (the maximum number of edges stored in memory). We also mention the number of graphs retained memory at the end of the stream. Note that $M_G$ grows proportional to $N$, and $M_Y$ is 32 times $M_X$, since projection vectors are integer vectors of the same length as sketches. Overall, \method's memory consumption for $N=12.5$M and $L=1000$ is as low as 240 {\sc mb}, which is comparable to the memory consumed by an average process on a commodity machine.

\begin{table}[h]
	\vspace{-0.15in}
	\caption{\method~memory use on \DC~($L = 1000$). $|G|$: final \# graphs in memory. Memory used by $M_G$:  graphs,  $M_Y$: projection vectors, \textbf{$M_X$}: sketches}
	\centering
	\begin{tabular}{rr||rrr}
		\toprule
		${N}$ & \textbf{$|G|$} & \textbf{$M_G$}  & \textbf{$M_Y$} & \textbf{$M_X$} \\
		\midrule
		1.25 {\sc m}		& 8 		& 23.84 {\sc mb}	& 31.25 {\sc kb}	& 0.98 	{\sc kb}	\\
		2.5 {\sc m}			& 29 		& 47.68 {\sc mb}	& 113.28 {\sc kb}	& 3.54 	{\sc kb}	\\
		3.75 {\sc m}		& 42 		& 71.52 {\sc mb}	& 164.06 {\sc kb}	& 5.13	{\sc kb}	\\
		6.25 {\sc m}		& 56 		& 119.20 {\sc mb}	& 218.75 {\sc kb}	& 6.84 	{\sc kb}	\\
		12.5 {\sc m}		& 125 		& 238.41 {\sc mb} 	& 488.28 {\sc kb}	& 15.26  {\sc kb}	\\
		\bottomrule
	\end{tabular}
	\label{tbl:efficiency}
	\vspace{-0.1in}
\end{table}

In summary, these results demonstrate the effectiveness as well as the time and memory-efficiency of \method.

%
%
%

%
%
%
%

\section{Related Work}
\label{sec:related}
Our work is related to a number of areas of study, including graph similarity, graph sketching and anomaly detection in streaming and typed graphs, which we detail further in this section. 

{\bf Graph similarity.}
There exists a large body of work on graph similarity, which can be used for various tasks including clustering and anomaly detection. Methods that require knowing the node correspondence between graphs are inapplicable to our scenario, as are methods that compute a vector of \emph{global} metrics \cite{conf/asonam/Berlingerio13} for each graph (such as the average properties across all nodes), since they cannot take into account the temporal ordering of edges.

\emph{Graph-edit-distance} (GED) \cite{bunke1983inexact} defines the dissimilarity between two graphs as the minimum total cost of operations required to make one graph isomorphic to the other. However, computing the GED requires finding an inexact matching between the two graphs that has minimum cost. This is known to be NP-hard and frequently addressed by local-search heuristics \cite{journals/datamine/Kostakis14} that provide no error bound.

\emph{Graph kernels} \cite{menchetti2005weighted, shervashidze2011weisfeiler, feragen2013scalable} decompose each graph into a set of local substructures and the similarity between two graphs is a function of the number of substructures they have in common. However, these methods require knowing a fixed universe of the substructures that constitute all the input graphs, which is unavailable in a streaming scenario.


{\bf Heterogeneous/typed graphs.}
An early method \cite{conf/kdd/NobleC03} used an information-theoretic approach to find anomalous node-typed graphs in a large static database. The method used the SUBDUE \cite{journals/expert/CookH00} system to first discover frequent substructures in the database in an offline fashion. The anomalies are then graphs containing only a few of the frequent substructures. Subsequent work \cite{conf/icdm/EberleH07} defined anomalies as graphs that were mostly similar to normative ones, but differing in a few GED operations. Frequent typed subgraphs were also leveraged as features to identify non-crashing software bugs from system execution flow graphs \cite{conf/sdm/LiuYYHY05}. Work also exists studying anomalous communities \cite{Gupta2014,Perozzi2016} and community anomalies \cite{conf/kdd/GaoLFWSH10,Perozzi2014} for attributed graphs. All of these approaches are designed for static graphs.


{\bf Streaming graphs.} {\sc{GMicro}} \cite{conf/sdm/AggarwalZY10} clustered untyped graph streams using a centroid-based approach and a distance function based on edge frequencies. It was extended to graphs with whole-graph-level attributes \cite{conf/sdm/YuZ13} and node-level attributes \cite{conf/sdm/McConville15}. {\sc{GOutlier}} \cite{conf/icde/AggarwalZY11} introduced structural reservoir sampling to maintain summaries of graph streams and detect anomalous graphs as those having unlikely edges. {\sc{Classy}} \cite{journals/datamine/Kostakis14} implemented a scalable distributed approach to clustering streams of call graphs by employing simulated annealing to approximate the GED between pairs of graphs, and GED lower bounds to prune away candidate clusters.

There also exist methods that detect changes in graphs that evolve through community evolution, by processing streaming edges to determine expanding and contracting communities \cite{conf/sdm/AggarwalY05}, and by applying information-theoretic \cite{conf/kdd/SunFPY07} and probabilistic \cite{conf/asunam/GuptaAHS11} approaches on graph snapshots to find time points of global community structure change. Methods also exist to find temporal patterns called graph evolution rules \cite{conf/pkdd/BerlingerioBBG09}, which are subgraphs with similar structure, types of nodes, and order of edges. Existing methods in this category are primarily concerned with untyped graphs.


{\bf Graph sketches and compact representations.} Graph ``skeletons'' \cite{conf/stoc/Karger94} were introduced to approximately solve a number of common graph-theoretic problems (e.g. global min-cut, max-flow) with provable error-bounds. Skeletons were  applied \cite{journals/pvldb/AggarwalXY09} to construct compressed representations of disk-resident graphs for efficiently approximating and answering minimum $s$-$t$ cut queries. Work on sketching graphs has primarily focused on constructing sketches that enable approximate solutions to specific graph problems such as finding the min-cut, testing reachability and finding the densest subgraph \cite{journals/sigmod/McGregor14}; the proposed sketches cannot be applied directly to detect graph-based anomalies.


\vspace{-0.05in}
\section{Conclusion}
\label{sec:conclude}
We have presented \method ~to cluster and detect anomalous heterogenous graphs originating from a stream of typed edges, in which new graphs emerge and existing graphs evolve as the stream progresses. We introduced representing heterogenous ordered graphs by shingling and devised \streamhash ~to maintain summaries of these representations online with constant-time updates and bounded memory consumption. Exploiting the mergeability of our summaries, we devised an online centroid-based clustering and anomaly detection scheme to rank incoming graphs by their anomalousness that obtains over 90\% average precision for the course of the stream. We showed that performance is sustained even under strict memory constraints, while being able to process over 100,000 edges per second.

While designed to detect APTs from system log streams, \method ~is applicable to other scenarios requiring scalable clustering and anomaly-ranking of typed graphs arriving in a stream of edges, for which no method currently exists. It has social media applications in event-detection using streams of sentences represented as syntax trees, or biochemical applications in detecting anomalous entities in streams of chemical compounds or protein structure elements.

%

\scriptsize{
\bibliographystyle{abbrv}
\bibliography{refs}
}

\end{document}